\documentclass[prl, reprint, superscriptaddress]{revtex4-2}
\usepackage{amsmath}
\usepackage{amssymb}
\usepackage{graphicx}
\usepackage[caption=false, position=top, singlelinecheck=off, justification=raggedright]{subfig}
\usepackage{multirow}
\usepackage{color}
\usepackage{pst-node}
\usepackage[unicode]{hyperref}
\hypersetup{
unicode=true,         	
colorlinks=true,      	
linkcolor=blue,		   	
citecolor=blue,       	
urlcolor=blue		   	
}

\begin{document}

\title{Dissipationless counterflow currents above $T_c$ in bilayer superconductors}

\author{Guido Homann}
\affiliation{Zentrum f\"ur Optische Quantentechnologien and Institut f\"ur Quantenphysik, 
	Universit\"at Hamburg, 22761 Hamburg, Germany}

\author{Marios H. Michael}
\affiliation{Max Planck Institute for the Structure and Dynamics of Matter, Luruper Chausse 149, 22761 Hamburg, Germany}
\email[Correspondence to: ]{marios.michael@mpsd.mpg.de}
\author{Jayson G. Cosme}
\affiliation{National Institute of Physics, University of the Philippines, Diliman, Quezon City 1101, Philippines}

\author{Ludwig Mathey}
\affiliation{Zentrum f\"ur Optische Quantentechnologien and Institut f\"ur Quantenphysik, 
	Universit\"at Hamburg, 22761 Hamburg, Germany}
\affiliation{The Hamburg Centre for Ultrafast Imaging, Luruper Chaussee 149, 22761 Hamburg, Germany}

\date{\today}
\begin{abstract}
    We report the existence of dissipationless currents in bilayer superconductors above the critical temperature $T_c$, assuming that the superconducting phase transition is dominated by phase fluctuations. Using a semiclassical $U(1)$ lattice gauge theory, we show that thermal fluctuations cause a transition from the superconducting state at low temperature to a resistive state above $T_c$, accompanied by the proliferation of unbound vortices. Remarkably, while the proliferation of vortex excitations causes dissipation of homogeneous in-plane currents, we find that counterflow currents, flowing in opposite direction within a bilayer, remain dissipationless. The presence of a dissipationless current channel above $T_c$ is attributed to the inhibition of vortex motion by local superconducting coherence within a single bilayer, in the presence of counterflow currents. Our theory presents a possible scenario for the pseudogap phase in bilayer cuprates.
\end{abstract}
\maketitle

\textit{Introduction} -- Underdoped cuprates exhibit two characteristic temperature scales. While these materials are superconducting only below the critical temperature $T_c$, they feature a gap-like suppression of the density of low-energy electronic states up to a significantly higher temperature $T^*$. The precise nature of this pseudogap regime is still under debate \cite{Norman2005, Keimer2015}. As the density of superconducting charge carriers is relatively small in underdoped cuprates, it was proposed that the breakdown of superconductivity at $T_c$ is dominated by phase fluctuations \cite{Emery1995}. This scenario, in which Cooper pairs exist up to $T^*$, is consistent with the similarity between the symmetries of the superconducting gap and the pseudogap \cite{Damascelli2003, Hashimoto2014}. Remarkably, measurements of the Nernst effect \cite{Xu2000, Wang2006, Daou2010, Cyr2018}, magnetization experiments \cite{Wang2005, Li2010} and optical spectroscopy \cite{Corson1999, Dubroka2011} indicate the existence of superconducting fluctuations well above $T_c$. Further evidence for superconducting fluctuations in the pseudogap regime is provided by pump-probe experiments
involving parametric amplification of Josephson plasmons in
YBa$_2$Cu$_3$O$_{7-\delta}$ (YBCO) \cite{Hu2014, Kaiser2014, VonHoegen2022, Michael2020}.

To contribute to the ongoing debate, we investigate the phenomenology of the pseudogap phase in bilayer superconductors, under the assumption that the transition is dominated by phase fluctuations while the pairing amplitude remains essentially constant up to temperatures close to $T^*$. We utilize a semiclassical $U(1)$ lattice gauge theory \cite{Homann2020, Homann2021, Homann2022} to simulate dynamics of the superconducting phase of a bilayer superconductor in the presence of thermal fluctuations. We find a crossover from an ordered state to a highly fluctuating state without global phase coherence at high temperatures, which we associate with the pseudogap phase. Our simulations show that in the pseudogap phase, long-range coherence is destroyed by the proliferation of vortex excitations, consistent with studies on the anisotropic $XY$ model \cite{Glazman1990, Janke1990, Minnhagen1991a, Minnhagen1991b, Fisher1991, Chattopadhyay1994}.

\begin{figure}[!b]
	\centering
	\includegraphics[scale=1]{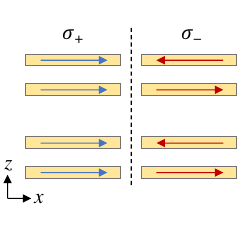}\llap{\parbox[b]{8.15cm}{{(a)}\\\rule{0ex}{3.4cm}}}
	\includegraphics[scale=1]{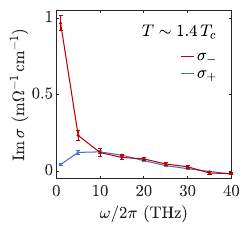}\llap{\parbox[b]{8.15cm}{{(b)}\\\rule{0ex}{3.4cm}}}
	\caption{Dissipationless counterflow in a bilayer superconductor. (a) Current configurations in the copper oxide layers characterized by the symmetric and the antisymmetric conductivity, respectively. (b) Imaginary part of the symmetric and the antisymmetric conductivity at $36~\mathrm{K} \sim 1.4 T_c$. The error bars indicate the standard errors of the ensemble averages.}
	\label{fig:fig1} 
\end{figure}

We find that the loss of long-range coherence is accompanied by a resistive transition where superconductivity is lost. However, our simulations show that short-range intrabilayer coherence persists far above $T_c$. A striking consequence of short-range intrabilayer superconducting correlations is the presence of a dissipationless counterflow current. To be precise, the in-plane conductivity of a bilayer superconductor can be divided into a symmetric and an anti-symmetric component as depicted in Fig.~\ref{fig:fig1}(a).
Our simulations, presented in Fig.\ref{fig:fig1}(b), show that the symmetric in-plane conductivity $\sigma_+ (\omega)$ no longer features a $1/\omega$ divergence at temperatures $T \gtrsim T_c$, signaling the emergence of a resistive state. In contrast, in-plane currents with opposite direction in the lower and upper layer flow without dissipation. This phenomenon manifests itself in a $1/\omega$ divergence of the antisymmetric conductivity $\sigma_- (\omega)$, as shown in Fig.~\ref{fig:fig1}(b). The effect that we present here is conceptually related to a variety of other phenomena, such as antisymmetric quasi-order in a bilayer of superfluids \cite{Mathey2008}, counterflow superfluidity in one-dimensional Bose mixtures \cite{Hu2009, Hu2011}, and Bose–Einstein condensation of excitons in bilayer electron systems \cite{Eisenstein1992, Tutuc2004, Eisenstein2004, Su2008}. We clarify that these phenomena are distinct from the normal-superfluid counterflow of the two-fluid model, which occurs within the superfluid phase\cite{Kobayashi19}.

\begin{figure}[!t]
	\centering
	\includegraphics[scale=1]{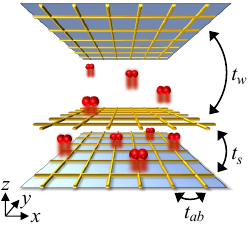}\llap{\parbox[b]{8.15cm}{{(a)}\\\rule{0ex}{3cm}}}
	\includegraphics[scale=1]{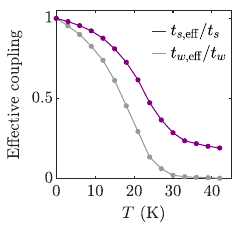}\llap{\parbox[b]{8.15cm}{{(b)}\\\rule{0ex}{3.4cm}}} \\
	\includegraphics[scale=1]{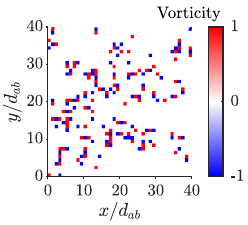}\llap{\parbox[b]{8.15cm}{{(c)}\\\rule{0ex}{3.45cm}}}
	\includegraphics[scale=1]{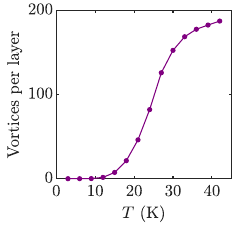}\llap{\parbox[b]{8.15cm}{{(d)}\\\rule{0ex}{3.45cm}}}
	\caption{Semiclassical simulation of a bilayer superconductor. (a) Schematic illustration of the lattice gauge model. (b) Temperature dependence of the effective interlayer tunneling coefficients. The intra-bilayer coherence remains non-zero above $T_c$. (c) Snapshot of the vorticity of the superconducting order parameter at $36~\mathrm{K} \sim 1.4 T_c$. (d) Temperature dependence of the number of vortices per layer. Vortices and antivortices contribute equally to this number. The vortex number rises sharply around $T_c$, suggesting that the transition is driven by vortex unbinding.}
	\label{fig:fig2} 
\end{figure}

\textit{Model} -- 
Following the Ginzburg-Landau theory of superconductivity \cite{Ginzburg1950}, we describe the superconducting state by a complex order parameter $\psi_{\mathbf{r}}= |\psi_{\mathbf{r}}| e^{i \phi_{\mathbf{r}}}$, which is discretized on a three-dimensional lattice with $\mathbf{r}$ being the lattice site. Each superconducting layer is represented by a square lattice as depicted in Fig.~\ref{fig:fig2}(a). The crystalline $c$~axis is oriented along the $z$ direction. Due to the Cooper pair charge of $-2e$, the order parameter is coupled to the electromagnetic field. We employ the Peierls substitution such that the electromagnetic vector potential $\mathbf{A}_{\mathbf{r}}$ enters the gauge-invariant phase differences between the lattice sites. The gauge-invariant phase differences, which are defined below, govern the nearest-neighbor tunneling of Cooper pairs.

In our model, we fit our parameters the bilayer cuprate YBCO\cite{ Hu2014, Kaiser2014, VonHoegen2022, Supp}. The interlayer distances are $d_s$ for intrabilayer (strong) junctions and $d_w$ for interbilayer (weak) junctions. Here we choose $d_s$ and $d_w$ such that the interlayer distances approximately reproduce the spacing of CuO$_2$ layers in the bilayer cuprate YBCO. The in-plane lattice constant $d_{ab}$ is introduced as a short-range cutoff below the in-plane coherence length. The tunneling coefficients are $t_{ab}$ for in-plane junctions, $t_s$ for intrabilayer junctions, and $t_w$ for interbilayer junctions.
We choose $t_s$ and $t_w$ such that the Josephson plasma frequencies of the simulated bilayer superconductor are comparable to those of YBCO: $\omega_{\mathrm{J1}}/2\pi \approx 1~\mathrm{THz}$ and $\omega_{\mathrm{J2}}/2\pi \approx 14~\mathrm{THz}$ . The in-plane tunneling coefficient $t_{ab}$ determines the in-plane plasma frequency, which we take to be $\omega_{ab}/2\pi \sim 75~\mathrm{THz}$ at zero temperature.

The Lagrangian and the equations of motion are presented in the Supplemental Material \cite{Supp}. We add Langevin noise and damping terms to the equations of motion and employ periodic boundary conditions. We then integrate the stochastic differential equations using Heun’s method with a step size of $\Delta t = 1.25~\mathrm{as}$.
In the following, we consider a bilayer superconductor with $N_z= 4$ layers and $N_{xy}= 40 \times 40$ sites per layer. The layers $n=1$ and $n=2$ form a bilayer, as well as the layers $n=3$ and $n=4$. The complete set of model parameters is specified in the Supplemental Material \cite{Supp}.

\begin{figure*}[!t]
	\centering
	\includegraphics[scale=1]{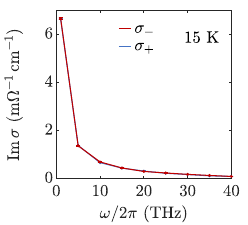}\llap{\parbox[b]{8.15cm}{{(a)}\\\rule{0ex}{3.4cm}}}
	\includegraphics[scale=1]{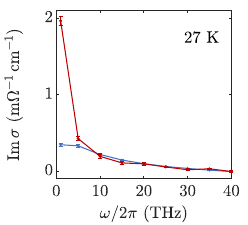}\llap{\parbox[b]{8.15cm}{{(b)}\\\rule{0ex}{3.4cm}}}
	\includegraphics[scale=1]{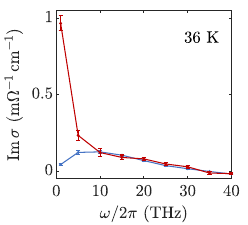}\llap{\parbox[b]{8.15cm}{{(c)}\\\rule{0ex}{3.4cm}}}
	\includegraphics[scale=1]{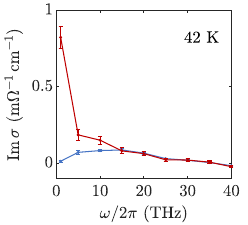}\llap{\parbox[b]{8.15cm}{{(d)}\\\rule{0ex}{3.4cm}}} \\
	\includegraphics[scale=1]{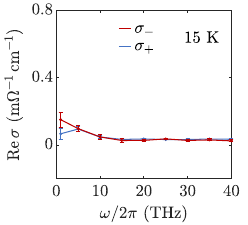}\llap{\parbox[b]{8.15cm}{{(e)}\\\rule{0ex}{3.4cm}}}
	\includegraphics[scale=1]{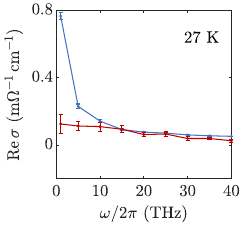}\llap{\parbox[b]{8.15cm}{{(f)}\\\rule{0ex}{3.4cm}}}
	\includegraphics[scale=1]{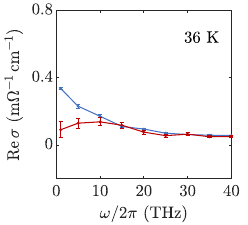}\llap{\parbox[b]{8.15cm}{{(g)}\\\rule{0ex}{3.4cm}}}
	\includegraphics[scale=1]{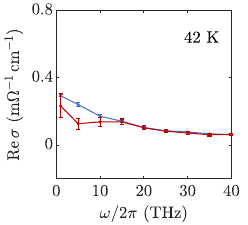}\llap{\parbox[b]{8.15cm}{{(h)}\\\rule{0ex}{3.4cm}}}
	\caption{Symmetric and antisymmetric conductivity at different temperatures. (a)--(d) Imaginary part. (e)--(h) Real part. The error bars indicate the standard errors of the ensemble averages. The crossover temperature is $T_c \sim 25~\mathrm{K}$.The anti-symmetric conductivity indicates superconductivity above $T_c$.}
	\label{fig:fig3}
\end{figure*}
\textit{Interlayer phase coherence} -- 
To characterize the coherence of the gauge-invariant intra- and interbilayer phase differences, we introduce effective interlayer tunneling coefficients. First, we define the unitless vector potential $a_{j,\mathbf{r}}= -2e d_{j,\mathbf{r}} A_{j,\mathbf{r}}/\hbar$ on the bond between the lattice site $\mathbf{r} \equiv (l,m,n)$ and its nearest neighbor in the $j \in \{x,y,z\}$ direction, where $d_{j,\mathbf{r}}$ denotes the length of the bond. The gauge-invariant intrabilayer phase differences between layers $n=1$ and $n=2$ are $\theta_{l,m}^s= \mathcal{P} (\phi_{l,m,1} - \phi_{l,m,2} + a_{l,m,1}^z)$, and the gauge-invariant interbilayer phase differences between layers $n=2$ and $n=3$ are $\theta_{l,m}^w= \mathcal{P} (\phi_{l,m,2} - \phi_{l,m,3} + a_{l,m,2}^z)$. Note that the gauge-invariant phase differences are mapped onto the interval $(-\pi,\pi]$ by the projection operator $\mathcal{P}(\cdot)$. In the presence of thermal fluctuations, we determine the effective interlayer tunneling coefficients
\begin{align}
	t_{s, \mathrm{eff}} &= t_s \left\langle \cos \theta_{l,m}^s \right\rangle , \label{eq:ts_eff} \\
	t_{w, \mathrm{eff}} &= t_w\left\langle \cos \theta_{l,m}^w \right\rangle . \label{eq:tw_eff}
\end{align}
The temperature dependence of the effective tunneling coefficients is shown in Fig.~\ref{fig:fig2}(b), where we average the cosine of the interlayer phase differences over the $xy$ plane, for a time interval of 2~ps, and an ensemble of 100 trajectories. The onset of strong phase fluctuations dramatically suppresses the interbilayer tunneling coefficient around a crossover temperature of $25~\mathrm{K}$, which we take to be the transition temperature $T_c$. While the advent of strong phase fluctuations suppresses also the intrabilayer coefficient $t_s$, it remains nonzero up to large temperatures. This indicates that while long-range order is lost across the transition, the pseudogap phase still retains strong local phase coherence within each bilayer. The consequences of phase fluctuations on the plasma resonances as well as the temperature dependence of the in-plane tunneling coefficient and the amplitude of the order parameter in our model are presented in the Supplemental Material  \cite{Supp}.

\textit{Vortices} -- 
To understand the microscopic nature of the phase transition, we turn our attention to the role of vortices in the pseudogap phase. In continuum theories, a vortex is defined through the phase winding of the order parameter along a closed path,
\begin{equation}
	\Phi= \oint \boldsymbol{\nabla} \phi \cdot \mathrm{d}\mathbf{r}= \oint \left(\boldsymbol{\nabla} \phi + \frac{2e}{\hbar} \mathbf{A} \right) \cdot \mathrm{d}\mathbf{r} \, - \, \oint \frac{2e}{\hbar} \mathbf{A} \cdot \mathrm{d}\mathbf{r} .
\end{equation}
In our simulation, we use the latter representation as it is based on quantities that directly enter the Lagrangian. We define the vorticity of a single plaquette in the $xy$ plane as
\begin{align} 
	\begin{split} \label{eq:vorticity}
		v_{l,m,n} ={}& \frac{1}{2\pi} \left(a_{l,m,n}^x + a_{l+1,m,n}^y - a_{l,m+1,n}^x - a_{l,m,n}^y \right) \\
		&- \frac{1}{2\pi} \left(\theta_{l,m,n}^x + \theta_{l+1,m,n}^y - \theta_{l,m+1,n}^x - \theta_{l,m,n}^y \right) ,
	\end{split}
\end{align}
where $\theta_{l,m,n}^x= \mathcal{P}(\phi_{l,m,n} - \phi_{l+1,m,n} + a_{l,m,n}^x)$ and $\theta_{l,m,n}^y= \mathcal{P}(\phi_{l,m,n} - \phi_{l,m+1,n} + a_{l,m,n}^y)$.
The vorticity can assume the values $-1$, 0, and $+1$. A vorticity of $+1$ corresponds to a vortex, while a vorticity of $-1$ corresponds to an antivortex. 

In Fig.~\ref{fig:fig2}(c), we show a snapshot of the vorticity in the lowest layer at a temperature of $36~\mathrm{K} \sim 1.4 T_c$. Even though most vortex and antivortex form pairs or clusters (seen as blue and red squares next to each other), we crucially also find isolated vortices and antivortices, indicating that the phase transition is driven by vortex-antivortex unbinding. In Figure~\ref{fig:fig2}(d), we plot the number of vortices per layer as a function of temperature. The number of vortices exhibits a rapid increase between 15~K and 30~K. Details of the behavior of vortices in the pseudogap phase are captured by computing in-plane and out-of plane correlations, which can be found in the Supplemental Material \cite{Supp}. In addition, the presence of vortices leads to a disordered Josephson intrabilayer potential. The strength and spectral behavior of the vortex-induced disorder is also presented in the Supplemental Material \cite{Supp}. Here we focus on the consequences of the phase transition on the conductivity.

\textit{In-plane conductivity} -- 
We separate the in-plane conductivity of a bilayer superconductor into a symmetric and an antisymmetric component as shown in Fig.~\ref{fig:fig1}(a). To calculate the two components of the conductivity, we introduce an oscillating symmetric (anti-symmetric) current, $J_{\pm}$. Once a steady state is reached, we compute $\sigma_{\pm}(\omega) = J_{\pm}(\omega) / E_{\pm}(\omega) $, where $E_{\pm}$ is the symmetric(anti-symmetric) electric field.  Details of the conductivity measurements are provided in the Supplemental Material \cite{Supp}.

The symmetric and the antisymmetric conductivity are shown for different temperatures in Fig.~\ref{fig:fig3}. At a temperature of $15~\mathrm{K} \sim 0.6 T_c$, $\sigma_+$ and $\sigma_-$ are in good agreement with each other. The imaginary part of $\sigma_+$ and $\sigma_-$ exhibits the characteristic $1/\omega$ behavior of a superconductor. While the real part of both conductivities is relatively flat for frequencies above 10~THz, it tends to slowly increase with decreasing frequency below 10~THz. The values of $\mathrm{Re} \, \sigma_+$ and $\mathrm{Re} \, \sigma_-$ at 1~THz bear some uncertainty due to slow numerical convergence.

At temperatures $T \gtrsim T_c$, the phase transition is accompanied by a dissipative transition, where the imaginary part of $\sigma_+$ no longer diverges as $1/\omega$. At temperatures close to $T_c$, the real part of $\sigma_+$ rises significantly at small frequencies. By contrast, the imaginary part of $\sigma_-$ exhibits a $1/\omega$ divergence up to temperatures well above $T_c$ while the real part has no significant temperature dependence. This manifestation of remnant superconductivity above $T_c$ is the key result of the present work.

\textit{Origin of dissipationless counterflow} -- The breakdown of the $1/\omega$ divergence of the imaginary part of $\sigma_+$ reveals a transition to a resistive state at $T_c$. This indicates an unbinding of planar vortex-antivortex pairs above $T_c$, similar to the resistive transition in superconducting thin films \cite{Halperin1979, Minnhagen1987}. The underlying mechanism of this transition is the following. In the presence of a current $\mathbf{J}$, a single vortex is exposed to a Magnus force $\mathbf{F}= \mathbf{J} \times \boldsymbol{\Phi_0}$, where $\mathbf{\Phi_0}$ has the magnitude of a flux quantum $\Phi_0= \pi \hbar/e$ and points in the direction of the magnetic field inside the vortex core. In the case of a DC current, unbound vortices and antivortices drift in opposite directions perpendicular to the current, dissipating energy. 

\begin{figure}[!t]
	\centering
	\includegraphics[scale=1]{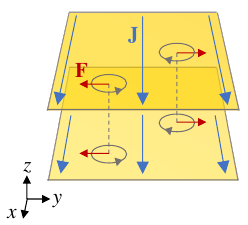}\llap{\parbox[b]{8cm}{{(a)}\\\rule{0ex}{3.4cm}}}
	\includegraphics[scale=1]{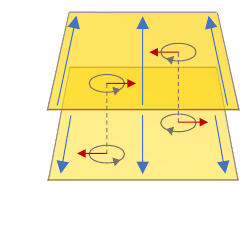}\llap{\parbox[b]{7.8cm}{{(b)}\\\rule{0ex}{3.4cm}}}
	\caption{Dynamics of intrabilayer vortex pairs in the presence of in-plane currents, sustaining dissipationless counterflow. (a) Vortex dynamics in the presence of unidirected in-plane currents. (b) Vortex dynamics in the presence of counterdirected in-plane currents. Vortices in different superconducting layers within the same bilayer are pinned relative to each other due to the residual superconducting coherence within a single bilayer.}
	\label{fig:fig4} 
\end{figure}

The simultaneous observation of dissipationless counterflow suggests that unbound vortex lines cut through an entire bilayer rather than just a single layer. Equivalently, a vortex in one layer of a bilayer is paired with a vortex of the same vorticity in the other layer of the same bilayer. In this scenario, the dissipation of currents in the two layers of the same bilayer is different depending on if the currents flow in the same or opposite directions. If the currents flow in the same direction, the Magnus force points in the same direction for all vortices in the two layers with the same vorticity. Thus, each vortex-vortex pair experiences a drift motion perpendicular to the current direction as depicted in Fig.~\ref{fig:fig4}(a). Analogously to the case of a thin film, the vortex motion dissipates energy, implying a nonzero resistivity. If the currents flow in opposite directions, however, the Magnus force points into opposite directions for the two vortices of each intrabilayer pair. The remnant intrabilayer coherence leads to an effective potential with a linear dependence on the pair size, acting as string tension against the flow of the two vortices away from each other in the presence of counterflow currents \cite{Girvin96}. Thus, the intrabilayer vortex pairs experience no net force as highlighted by Fig.~\ref{fig:fig4}(b). Since the vortices do not move in this case, the flow of counterdirected currents is dissipationless, consistent with the observation of a $1/\omega$ divergence of the imaginary part of  $\sigma_-$ above $T_c$. 

We note that the previous paragraph provides only a simplified description of the vortex dynamics in the presence of in-plane currents. In fact, the vortex dynamics is very complicated due to the high density of vortices in the layers and fast creation/annihilation processes; see Supplemental Material \cite{Supp}.
Nonetheless, the scenario of intrabilayer vortex-vortex pairs that are essentially unbound from any antivortices is supported by several correlation functions \cite{Supp}. 
In the Supplemental Material, we also calculate conductivities at finite momentum along the $z$ direction \cite{Supp}. The results are momentum independent, which corroborates the picture of decoupled bilayers, where vortex lines between different bilayers are uncorrelated.

\textit{Conclusion} -- We have discovered the existence of dissipationless counterflow currents in a $U(1)$ gauge-invariant model for bilayer superconductors. Experimental verification of the existence of dissipationless counterflow in bilayer cuprates would provide smoking gun evidence that the pseudogap phase corresponds to phase-fluctuating superconductivity with strong intrabilayer superconducting correlations up to high temperatures. We expect counterflow currents to appear when a magnetic field is applied in parallel to the layers, giving rise to a diamagnetic response \cite{Fan16}. The results presented here open up interesting research questions about the full range of consequences of such dissipationless currents and whether they can be technologically exploited.

\begin{acknowledgments}
	We would like to acknowledge useful discussions with Patrick A. Lee and Eugene Demler.
	This work is supported by the Deutsche Forschungsgemeinschaft (DFG) in the framework of SFB~925, Project No.~170620586, and the Cluster of Excellence ``Advanced Imaging of Matter" (EXC~2056), Project No.~390715994. M.H.M. is grateful for the support from the Alexander von Humboldt foundation and the hospitality of the Max Planck Institute for the Structure and Dynamics of Matter.  J.G.C. is funded by the UP System Balik PhD Program (OVPAA-BPhD-2021-04).
	
\end{acknowledgments}

\bibliography{biblio}

\providecommand{\noopsort}[1]{}\providecommand{\singleletter}[1]{#1}%
\begin{thebibliography}{40}%
\makeatletter
\providecommand \@ifxundefined [1]{%
 \@ifx{#1\undefined}
}%
\providecommand \@ifnum [1]{%
 \ifnum #1\expandafter \@firstoftwo
 \else \expandafter \@secondoftwo
 \fi
}%
\providecommand \@ifx [1]{%
 \ifx #1\expandafter \@firstoftwo
 \else \expandafter \@secondoftwo
 \fi
}%
\providecommand \natexlab [1]{#1}%
\providecommand \enquote  [1]{``#1''}%
\providecommand \bibnamefont  [1]{#1}%
\providecommand \bibfnamefont [1]{#1}%
\providecommand \citenamefont [1]{#1}%
\providecommand \href@noop [0]{\@secondoftwo}%
\providecommand \href [0]{\begingroup \@sanitize@url \@href}%
\providecommand \@href[1]{\@@startlink{#1}\@@href}%
\providecommand \@@href[1]{\endgroup#1\@@endlink}%
\providecommand \@sanitize@url [0]{\catcode `\\12\catcode `\$12\catcode
  `\&12\catcode `\#12\catcode `\^12\catcode `\_12\catcode `\%12\relax}%
\providecommand \@@startlink[1]{}%
\providecommand \@@endlink[0]{}%
\providecommand \url  [0]{\begingroup\@sanitize@url \@url }%
\providecommand \@url [1]{\endgroup\@href {#1}{\urlprefix }}%
\providecommand \urlprefix  [0]{URL }%
\providecommand \Eprint [0]{\href }%
\providecommand \doibase [0]{https://doi.org/}%
\providecommand \selectlanguage [0]{\@gobble}%
\providecommand \bibinfo  [0]{\@secondoftwo}%
\providecommand \bibfield  [0]{\@secondoftwo}%
\providecommand \translation [1]{[#1]}%
\providecommand \BibitemOpen [0]{}%
\providecommand \bibitemStop [0]{}%
\providecommand \bibitemNoStop [0]{.\EOS\space}%
\providecommand \EOS [0]{\spacefactor3000\relax}%
\providecommand \BibitemShut  [1]{\csname bibitem#1\endcsname}%
\let\auto@bib@innerbib\@empty
\bibitem [{\citenamefont {Norman}\ \emph {et~al.}(2005)\citenamefont {Norman},
  \citenamefont {Pines},\ and\ \citenamefont {Kalli}}]{Norman2005}%
  \BibitemOpen
  \bibfield  {author} {\bibinfo {author} {\bibfnamefont {M.~R.}\ \bibnamefont
  {Norman}}, \bibinfo {author} {\bibfnamefont {D.}~\bibnamefont {Pines}},\ and\
  \bibinfo {author} {\bibfnamefont {C.}~\bibnamefont {Kalli}},\ }\bibfield
  {title} {\bibinfo {title} {The pseudogap: friend or foe of high {$T_c$}?},\
  }\href {https://doi.org/10.1080/00018730500459906} {\bibfield  {journal}
  {\bibinfo  {journal} {Adv. Phys.}\ }\textbf {\bibinfo {volume} {54}},\
  \bibinfo {pages} {715} (\bibinfo {year} {2005})}\BibitemShut {NoStop}%
\bibitem [{\citenamefont {{Keimer}}\ \emph {et~al.}(2015)\citenamefont
  {{Keimer}}, \citenamefont {{Kivelson}}, \citenamefont {{Norman}},
  \citenamefont {{Uchida}},\ and\ \citenamefont {{Zaanen}}}]{Keimer2015}%
  \BibitemOpen
  \bibfield  {author} {\bibinfo {author} {\bibfnamefont {B.}~\bibnamefont
  {{Keimer}}}, \bibinfo {author} {\bibfnamefont {S.~A.}\ \bibnamefont
  {{Kivelson}}}, \bibinfo {author} {\bibfnamefont {M.~R.}\ \bibnamefont
  {{Norman}}}, \bibinfo {author} {\bibfnamefont {S.}~\bibnamefont {{Uchida}}},\
  and\ \bibinfo {author} {\bibfnamefont {J.}~\bibnamefont {{Zaanen}}},\
  }\bibfield  {title} {\bibinfo {title} {From quantum matter to
  high-temperature superconductivity in copper oxides},\ }\href
  {https://doi.org/10.1038/nature14165} {\bibfield  {journal} {\bibinfo
  {journal} {Nature}\ }\textbf {\bibinfo {volume} {518}},\ \bibinfo {pages}
  {179} (\bibinfo {year} {2015})}\BibitemShut {NoStop}%
\bibitem [{\citenamefont {{Emery}}\ and\ \citenamefont
  {{Kivelson}}(1995)}]{Emery1995}%
  \BibitemOpen
  \bibfield  {author} {\bibinfo {author} {\bibfnamefont {V.~J.}\ \bibnamefont
  {{Emery}}}\ and\ \bibinfo {author} {\bibfnamefont {S.~A.}\ \bibnamefont
  {{Kivelson}}},\ }\bibfield  {title} {\bibinfo {title} {Importance of phase
  fluctuations in superconductors with small superfluid density},\ }\href
  {https://doi.org/10.1038/374434a0} {\bibfield  {journal} {\bibinfo  {journal}
  {Nature}\ }\textbf {\bibinfo {volume} {374}},\ \bibinfo {pages} {434}
  (\bibinfo {year} {1995})}\BibitemShut {NoStop}%
\bibitem [{\citenamefont {Damascelli}\ \emph {et~al.}(2003)\citenamefont
  {Damascelli}, \citenamefont {Hussain},\ and\ \citenamefont
  {Shen}}]{Damascelli2003}%
  \BibitemOpen
  \bibfield  {author} {\bibinfo {author} {\bibfnamefont {A.}~\bibnamefont
  {Damascelli}}, \bibinfo {author} {\bibfnamefont {Z.}~\bibnamefont
  {Hussain}},\ and\ \bibinfo {author} {\bibfnamefont {Z.-X.}\ \bibnamefont
  {Shen}},\ }\bibfield  {title} {\bibinfo {title} {Angle-resolved photoemission
  studies of the cuprate superconductors},\ }\href
  {https://doi.org/10.1103/RevModPhys.75.473} {\bibfield  {journal} {\bibinfo
  {journal} {Rev. Mod. Phys.}\ }\textbf {\bibinfo {volume} {75}},\ \bibinfo
  {pages} {473} (\bibinfo {year} {2003})}\BibitemShut {NoStop}%
\bibitem [{\citenamefont {{Hashimoto}}\ \emph {et~al.}(2014)\citenamefont
  {{Hashimoto}}, \citenamefont {{Vishik}}, \citenamefont {{He}}, \citenamefont
  {{Devereaux}},\ and\ \citenamefont {{Shen}}}]{Hashimoto2014}%
  \BibitemOpen
  \bibfield  {author} {\bibinfo {author} {\bibfnamefont {M.}~\bibnamefont
  {{Hashimoto}}}, \bibinfo {author} {\bibfnamefont {I.~M.}\ \bibnamefont
  {{Vishik}}}, \bibinfo {author} {\bibfnamefont {R.-H.}\ \bibnamefont {{He}}},
  \bibinfo {author} {\bibfnamefont {T.~P.}\ \bibnamefont {{Devereaux}}},\ and\
  \bibinfo {author} {\bibfnamefont {Z.-X.}\ \bibnamefont {{Shen}}},\ }\bibfield
   {title} {\bibinfo {title} {Energy gaps in high-transition-temperature
  cuprate superconductors},\ }\href {https://doi.org/10.1038/nphys3009}
  {\bibfield  {journal} {\bibinfo  {journal} {Nat. Phys.}\ }\textbf {\bibinfo
  {volume} {10}},\ \bibinfo {pages} {483} (\bibinfo {year} {2014})}\BibitemShut
  {NoStop}%
\bibitem [{\citenamefont {Xu}\ \emph {et~al.}(2000)\citenamefont {Xu},
  \citenamefont {Ong}, \citenamefont {Wang}, \citenamefont {Kakeshita},\ and\
  \citenamefont {Uchida}}]{Xu2000}%
  \BibitemOpen
  \bibfield  {author} {\bibinfo {author} {\bibfnamefont {Z.~A.}\ \bibnamefont
  {Xu}}, \bibinfo {author} {\bibfnamefont {N.~P.}\ \bibnamefont {Ong}},
  \bibinfo {author} {\bibfnamefont {Y.}~\bibnamefont {Wang}}, \bibinfo {author}
  {\bibfnamefont {T.}~\bibnamefont {Kakeshita}},\ and\ \bibinfo {author}
  {\bibfnamefont {S.}~\bibnamefont {Uchida}},\ }\bibfield  {title} {\bibinfo
  {title} {Vortex-like excitations and the onset of superconducting phase
  fluctuation in underdoped {La$_{2-x}$Sr$_x$CuO$_4$}},\ }\href
  {https://doi.org/10.1038/35020016} {\bibfield  {journal} {\bibinfo  {journal}
  {Nature}\ }\textbf {\bibinfo {volume} {406}},\ \bibinfo {pages} {486}
  (\bibinfo {year} {2000})}\BibitemShut {NoStop}%
\bibitem [{\citenamefont {Wang}\ \emph {et~al.}(2006)\citenamefont {Wang},
  \citenamefont {Li},\ and\ \citenamefont {Ong}}]{Wang2006}%
  \BibitemOpen
  \bibfield  {author} {\bibinfo {author} {\bibfnamefont {Y.}~\bibnamefont
  {Wang}}, \bibinfo {author} {\bibfnamefont {L.}~\bibnamefont {Li}},\ and\
  \bibinfo {author} {\bibfnamefont {N.~P.}\ \bibnamefont {Ong}},\ }\bibfield
  {title} {\bibinfo {title} {Nernst effect in high-${T}_{c}$ superconductors},\
  }\href {https://doi.org/10.1103/PhysRevB.73.024510} {\bibfield  {journal}
  {\bibinfo  {journal} {Phys. Rev. B}\ }\textbf {\bibinfo {volume} {73}},\
  \bibinfo {pages} {024510} (\bibinfo {year} {2006})}\BibitemShut {NoStop}%
\bibitem [{\citenamefont {{Daou}}\ \emph {et~al.}(2010)\citenamefont {{Daou}},
  \citenamefont {{Chang}}, \citenamefont {{LeBoeuf}}, \citenamefont
  {{Cyr-Choini{\`e}re}}, \citenamefont {{Lalibert{\'e}}}, \citenamefont
  {{Doiron-Leyraud}}, \citenamefont {{Ramshaw}}, \citenamefont {{Liang}},
  \citenamefont {{Bonn}}, \citenamefont {{Hardy}},\ and\ \citenamefont
  {{Taillefer}}}]{Daou2010}%
  \BibitemOpen
  \bibfield  {author} {\bibinfo {author} {\bibfnamefont {R.}~\bibnamefont
  {{Daou}}}, \bibinfo {author} {\bibfnamefont {J.}~\bibnamefont {{Chang}}},
  \bibinfo {author} {\bibfnamefont {D.}~\bibnamefont {{LeBoeuf}}}, \bibinfo
  {author} {\bibfnamefont {O.}~\bibnamefont {{Cyr-Choini{\`e}re}}}, \bibinfo
  {author} {\bibfnamefont {F.}~\bibnamefont {{Lalibert{\'e}}}}, \bibinfo
  {author} {\bibfnamefont {N.}~\bibnamefont {{Doiron-Leyraud}}}, \bibinfo
  {author} {\bibfnamefont {B.~J.}\ \bibnamefont {{Ramshaw}}}, \bibinfo {author}
  {\bibfnamefont {R.}~\bibnamefont {{Liang}}}, \bibinfo {author} {\bibfnamefont
  {D.~A.}\ \bibnamefont {{Bonn}}}, \bibinfo {author} {\bibfnamefont {W.~N.}\
  \bibnamefont {{Hardy}}},\ and\ \bibinfo {author} {\bibfnamefont
  {L.}~\bibnamefont {{Taillefer}}},\ }\bibfield  {title} {\bibinfo {title}
  {Broken rotational symmetry in the pseudogap phase of a {high-T$_{c}$}
  superconductor},\ }\href {https://doi.org/10.1038/nature08716} {\bibfield
  {journal} {\bibinfo  {journal} {Nature}\ }\textbf {\bibinfo {volume} {463}},\
  \bibinfo {pages} {519} (\bibinfo {year} {2010})}\BibitemShut {NoStop}%
\bibitem [{\citenamefont {Cyr-Choini\`ere}\ \emph {et~al.}(2018)\citenamefont
  {Cyr-Choini\`ere}, \citenamefont {Daou}, \citenamefont {Lalibert\'e},
  \citenamefont {Collignon}, \citenamefont {Badoux}, \citenamefont {LeBoeuf},
  \citenamefont {Chang}, \citenamefont {Ramshaw}, \citenamefont {Bonn},
  \citenamefont {Hardy}, \citenamefont {Liang}, \citenamefont {Yan},
  \citenamefont {Cheng}, \citenamefont {Zhou}, \citenamefont {Goodenough},
  \citenamefont {Pyon}, \citenamefont {Takayama}, \citenamefont {Takagi},
  \citenamefont {Doiron-Leyraud},\ and\ \citenamefont {Taillefer}}]{Cyr2018}%
  \BibitemOpen
  \bibfield  {author} {\bibinfo {author} {\bibfnamefont {O.}~\bibnamefont
  {Cyr-Choini\`ere}}, \bibinfo {author} {\bibfnamefont {R.}~\bibnamefont
  {Daou}}, \bibinfo {author} {\bibfnamefont {F.}~\bibnamefont {Lalibert\'e}},
  \bibinfo {author} {\bibfnamefont {C.}~\bibnamefont {Collignon}}, \bibinfo
  {author} {\bibfnamefont {S.}~\bibnamefont {Badoux}}, \bibinfo {author}
  {\bibfnamefont {D.}~\bibnamefont {LeBoeuf}}, \bibinfo {author} {\bibfnamefont
  {J.}~\bibnamefont {Chang}}, \bibinfo {author} {\bibfnamefont {B.~J.}\
  \bibnamefont {Ramshaw}}, \bibinfo {author} {\bibfnamefont {D.~A.}\
  \bibnamefont {Bonn}}, \bibinfo {author} {\bibfnamefont {W.~N.}\ \bibnamefont
  {Hardy}}, \bibinfo {author} {\bibfnamefont {R.}~\bibnamefont {Liang}},
  \bibinfo {author} {\bibfnamefont {J.-Q.}\ \bibnamefont {Yan}}, \bibinfo
  {author} {\bibfnamefont {J.-G.}\ \bibnamefont {Cheng}}, \bibinfo {author}
  {\bibfnamefont {J.-S.}\ \bibnamefont {Zhou}}, \bibinfo {author}
  {\bibfnamefont {J.~B.}\ \bibnamefont {Goodenough}}, \bibinfo {author}
  {\bibfnamefont {S.}~\bibnamefont {Pyon}}, \bibinfo {author} {\bibfnamefont
  {T.}~\bibnamefont {Takayama}}, \bibinfo {author} {\bibfnamefont
  {H.}~\bibnamefont {Takagi}}, \bibinfo {author} {\bibfnamefont
  {N.}~\bibnamefont {Doiron-Leyraud}},\ and\ \bibinfo {author} {\bibfnamefont
  {L.}~\bibnamefont {Taillefer}},\ }\bibfield  {title} {\bibinfo {title}
  {Pseudogap temperature ${T}^{*}$ of cuprate superconductors from the {Nernst}
  effect},\ }\href {https://doi.org/10.1103/PhysRevB.97.064502} {\bibfield
  {journal} {\bibinfo  {journal} {Phys. Rev. B}\ }\textbf {\bibinfo {volume}
  {97}},\ \bibinfo {pages} {064502} (\bibinfo {year} {2018})}\BibitemShut
  {NoStop}%
\bibitem [{\citenamefont {Wang}\ \emph {et~al.}(2005)\citenamefont {Wang},
  \citenamefont {Li}, \citenamefont {Naughton}, \citenamefont {Gu},
  \citenamefont {Uchida},\ and\ \citenamefont {Ong}}]{Wang2005}%
  \BibitemOpen
  \bibfield  {author} {\bibinfo {author} {\bibfnamefont {Y.}~\bibnamefont
  {Wang}}, \bibinfo {author} {\bibfnamefont {L.}~\bibnamefont {Li}}, \bibinfo
  {author} {\bibfnamefont {M.~J.}\ \bibnamefont {Naughton}}, \bibinfo {author}
  {\bibfnamefont {G.~D.}\ \bibnamefont {Gu}}, \bibinfo {author} {\bibfnamefont
  {S.}~\bibnamefont {Uchida}},\ and\ \bibinfo {author} {\bibfnamefont {N.~P.}\
  \bibnamefont {Ong}},\ }\bibfield  {title} {\bibinfo {title} {Field-enhanced
  diamagnetism in the pseudogap state of the cuprate
  {${\mathrm{Bi}}_{2}{\mathrm{Sr}}_{2}\mathrm{Ca}{\mathrm{Cu}}_{2}{\mathrm{O}}_{8+\ensuremath{\delta}}$}
  superconductor in an intense magnetic field},\ }\href
  {https://doi.org/10.1103/PhysRevLett.95.247002} {\bibfield  {journal}
  {\bibinfo  {journal} {Phys. Rev. Lett.}\ }\textbf {\bibinfo {volume} {95}},\
  \bibinfo {pages} {247002} (\bibinfo {year} {2005})}\BibitemShut {NoStop}%
\bibitem [{\citenamefont {Li}\ \emph {et~al.}(2010)\citenamefont {Li},
  \citenamefont {Wang}, \citenamefont {Komiya}, \citenamefont {Ono},
  \citenamefont {Ando}, \citenamefont {Gu},\ and\ \citenamefont
  {Ong}}]{Li2010}%
  \BibitemOpen
  \bibfield  {author} {\bibinfo {author} {\bibfnamefont {L.}~\bibnamefont
  {Li}}, \bibinfo {author} {\bibfnamefont {Y.}~\bibnamefont {Wang}}, \bibinfo
  {author} {\bibfnamefont {S.}~\bibnamefont {Komiya}}, \bibinfo {author}
  {\bibfnamefont {S.}~\bibnamefont {Ono}}, \bibinfo {author} {\bibfnamefont
  {Y.}~\bibnamefont {Ando}}, \bibinfo {author} {\bibfnamefont {G.~D.}\
  \bibnamefont {Gu}},\ and\ \bibinfo {author} {\bibfnamefont {N.~P.}\
  \bibnamefont {Ong}},\ }\bibfield  {title} {\bibinfo {title} {Diamagnetism and
  {Cooper} pairing above ${T}_{c}$ in cuprates},\ }\href
  {https://doi.org/10.1103/PhysRevB.81.054510} {\bibfield  {journal} {\bibinfo
  {journal} {Phys. Rev. B}\ }\textbf {\bibinfo {volume} {81}},\ \bibinfo
  {pages} {054510} (\bibinfo {year} {2010})}\BibitemShut {NoStop}%
\bibitem [{\citenamefont {{Corson}}\ \emph {et~al.}(1999)\citenamefont
  {{Corson}}, \citenamefont {{Mallozzi}}, \citenamefont {{Orenstein}},
  \citenamefont {{Eckstein}},\ and\ \citenamefont {{Bozovic}}}]{Corson1999}%
  \BibitemOpen
  \bibfield  {author} {\bibinfo {author} {\bibfnamefont {J.}~\bibnamefont
  {{Corson}}}, \bibinfo {author} {\bibfnamefont {R.}~\bibnamefont
  {{Mallozzi}}}, \bibinfo {author} {\bibfnamefont {J.}~\bibnamefont
  {{Orenstein}}}, \bibinfo {author} {\bibfnamefont {J.~N.}\ \bibnamefont
  {{Eckstein}}},\ and\ \bibinfo {author} {\bibfnamefont {I.}~\bibnamefont
  {{Bozovic}}},\ }\bibfield  {title} {\bibinfo {title} {Vanishing of phase
  coherence in underdoped
  {Bi$_{2}$Sr$_{2}$CaCu$_{2}$O$_{8+{\ensuremath{\delta}}}$}},\ }\href
  {https://doi.org/10.1038/18402} {\bibfield  {journal} {\bibinfo  {journal}
  {Nature}\ }\textbf {\bibinfo {volume} {398}},\ \bibinfo {pages} {221}
  (\bibinfo {year} {1999})}\BibitemShut {NoStop}%
\bibitem [{\citenamefont {Dubroka}\ \emph {et~al.}(2011)\citenamefont
  {Dubroka}, \citenamefont {R\"ossle}, \citenamefont {Kim}, \citenamefont
  {Malik}, \citenamefont {Munzar}, \citenamefont {Basov}, \citenamefont
  {Schafgans}, \citenamefont {Moon}, \citenamefont {Lin}, \citenamefont {Haug},
  \citenamefont {Hinkov}, \citenamefont {Keimer}, \citenamefont {Wolf},
  \citenamefont {Storey}, \citenamefont {Tallon},\ and\ \citenamefont
  {Bernhard}}]{Dubroka2011}%
  \BibitemOpen
  \bibfield  {author} {\bibinfo {author} {\bibfnamefont {A.}~\bibnamefont
  {Dubroka}}, \bibinfo {author} {\bibfnamefont {M.}~\bibnamefont {R\"ossle}},
  \bibinfo {author} {\bibfnamefont {K.~W.}\ \bibnamefont {Kim}}, \bibinfo
  {author} {\bibfnamefont {V.~K.}\ \bibnamefont {Malik}}, \bibinfo {author}
  {\bibfnamefont {D.}~\bibnamefont {Munzar}}, \bibinfo {author} {\bibfnamefont
  {D.~N.}\ \bibnamefont {Basov}}, \bibinfo {author} {\bibfnamefont {A.~A.}\
  \bibnamefont {Schafgans}}, \bibinfo {author} {\bibfnamefont {S.~J.}\
  \bibnamefont {Moon}}, \bibinfo {author} {\bibfnamefont {C.~T.}\ \bibnamefont
  {Lin}}, \bibinfo {author} {\bibfnamefont {D.}~\bibnamefont {Haug}}, \bibinfo
  {author} {\bibfnamefont {V.}~\bibnamefont {Hinkov}}, \bibinfo {author}
  {\bibfnamefont {B.}~\bibnamefont {Keimer}}, \bibinfo {author} {\bibfnamefont
  {T.}~\bibnamefont {Wolf}}, \bibinfo {author} {\bibfnamefont {J.~G.}\
  \bibnamefont {Storey}}, \bibinfo {author} {\bibfnamefont {J.~L.}\
  \bibnamefont {Tallon}},\ and\ \bibinfo {author} {\bibfnamefont
  {C.}~\bibnamefont {Bernhard}},\ }\bibfield  {title} {\bibinfo {title}
  {Evidence of a precursor superconducting phase at temperatures as high as
  {180 K} in {RBa$_{2}$Cu$_{3}$O$_{7\ensuremath{-}\ensuremath{\delta}}$
  ($R=Y,Gd,Eu$)} superconducting crystals from infrared spectroscopy},\ }\href
  {https://doi.org/10.1103/PhysRevLett.106.047006} {\bibfield  {journal}
  {\bibinfo  {journal} {Phys. Rev. Lett.}\ }\textbf {\bibinfo {volume} {106}},\
  \bibinfo {pages} {047006} (\bibinfo {year} {2011})}\BibitemShut {NoStop}%
\bibitem [{\citenamefont {{Hu}}\ \emph {et~al.}(2014)\citenamefont {{Hu}},
  \citenamefont {{Kaiser}}, \citenamefont {{Nicoletti}}, \citenamefont
  {{Hunt}}, \citenamefont {{Gierz}}, \citenamefont {{Hoffmann}}, \citenamefont
  {{Le Tacon}}, \citenamefont {{Loew}}, \citenamefont {{Keimer}},\ and\
  \citenamefont {{Cavalleri}}}]{Hu2014}%
  \BibitemOpen
  \bibfield  {author} {\bibinfo {author} {\bibfnamefont {W.}~\bibnamefont
  {{Hu}}}, \bibinfo {author} {\bibfnamefont {S.}~\bibnamefont {{Kaiser}}},
  \bibinfo {author} {\bibfnamefont {D.}~\bibnamefont {{Nicoletti}}}, \bibinfo
  {author} {\bibfnamefont {C.~R.}\ \bibnamefont {{Hunt}}}, \bibinfo {author}
  {\bibfnamefont {I.}~\bibnamefont {{Gierz}}}, \bibinfo {author} {\bibfnamefont
  {M.~C.}\ \bibnamefont {{Hoffmann}}}, \bibinfo {author} {\bibfnamefont
  {M.}~\bibnamefont {{Le Tacon}}}, \bibinfo {author} {\bibfnamefont
  {T.}~\bibnamefont {{Loew}}}, \bibinfo {author} {\bibfnamefont
  {B.}~\bibnamefont {{Keimer}}},\ and\ \bibinfo {author} {\bibfnamefont
  {A.}~\bibnamefont {{Cavalleri}}},\ }\bibfield  {title} {\bibinfo {title}
  {Optically enhanced coherent transport in {YBa$_{2}$Cu$_{3}$O$_{6.5}$} by
  ultrafast redistribution of interlayer coupling},\ }\href
  {https://doi.org/10.1038/nmat3963} {\bibfield  {journal} {\bibinfo  {journal}
  {Nat. Mater.}\ }\textbf {\bibinfo {volume} {13}},\ \bibinfo {pages} {705}
  (\bibinfo {year} {2014})}\BibitemShut {NoStop}%
\bibitem [{\citenamefont {Kaiser}\ \emph {et~al.}(2014)\citenamefont {Kaiser},
  \citenamefont {Hunt}, \citenamefont {Nicoletti}, \citenamefont {Hu},
  \citenamefont {Gierz}, \citenamefont {Liu}, \citenamefont {Le~Tacon},
  \citenamefont {Loew}, \citenamefont {Haug}, \citenamefont {Keimer},\ and\
  \citenamefont {Cavalleri}}]{Kaiser2014}%
  \BibitemOpen
  \bibfield  {author} {\bibinfo {author} {\bibfnamefont {S.}~\bibnamefont
  {Kaiser}}, \bibinfo {author} {\bibfnamefont {C.~R.}\ \bibnamefont {Hunt}},
  \bibinfo {author} {\bibfnamefont {D.}~\bibnamefont {Nicoletti}}, \bibinfo
  {author} {\bibfnamefont {W.}~\bibnamefont {Hu}}, \bibinfo {author}
  {\bibfnamefont {I.}~\bibnamefont {Gierz}}, \bibinfo {author} {\bibfnamefont
  {H.~Y.}\ \bibnamefont {Liu}}, \bibinfo {author} {\bibfnamefont
  {M.}~\bibnamefont {Le~Tacon}}, \bibinfo {author} {\bibfnamefont
  {T.}~\bibnamefont {Loew}}, \bibinfo {author} {\bibfnamefont {D.}~\bibnamefont
  {Haug}}, \bibinfo {author} {\bibfnamefont {B.}~\bibnamefont {Keimer}},\ and\
  \bibinfo {author} {\bibfnamefont {A.}~\bibnamefont {Cavalleri}},\ }\bibfield
  {title} {\bibinfo {title} {Optically induced coherent transport far above
  ${T}_{c}$ in underdoped
  {${\mathrm{YBa}}_{2}{\mathrm{Cu}}_{3}{\mathrm{O}}_{6+\ensuremath{\delta}}$}},\
  }\href {https://doi.org/10.1103/PhysRevB.89.184516} {\bibfield  {journal}
  {\bibinfo  {journal} {Phys. Rev. B}\ }\textbf {\bibinfo {volume} {89}},\
  \bibinfo {pages} {184516} (\bibinfo {year} {2014})}\BibitemShut {NoStop}%
\bibitem [{\citenamefont {von Hoegen}\ \emph {et~al.}(2022)\citenamefont {von
  Hoegen}, \citenamefont {Fechner}, \citenamefont {F\"orst}, \citenamefont
  {Taherian}, \citenamefont {Rowe}, \citenamefont {Ribak}, \citenamefont
  {Porras}, \citenamefont {Keimer}, \citenamefont {Michael}, \citenamefont
  {Demler},\ and\ \citenamefont {Cavalleri}}]{VonHoegen2022}%
  \BibitemOpen
  \bibfield  {author} {\bibinfo {author} {\bibfnamefont {A.}~\bibnamefont {von
  Hoegen}}, \bibinfo {author} {\bibfnamefont {M.}~\bibnamefont {Fechner}},
  \bibinfo {author} {\bibfnamefont {M.}~\bibnamefont {F\"orst}}, \bibinfo
  {author} {\bibfnamefont {N.}~\bibnamefont {Taherian}}, \bibinfo {author}
  {\bibfnamefont {E.}~\bibnamefont {Rowe}}, \bibinfo {author} {\bibfnamefont
  {A.}~\bibnamefont {Ribak}}, \bibinfo {author} {\bibfnamefont
  {J.}~\bibnamefont {Porras}}, \bibinfo {author} {\bibfnamefont
  {B.}~\bibnamefont {Keimer}}, \bibinfo {author} {\bibfnamefont
  {M.}~\bibnamefont {Michael}}, \bibinfo {author} {\bibfnamefont
  {E.}~\bibnamefont {Demler}},\ and\ \bibinfo {author} {\bibfnamefont
  {A.}~\bibnamefont {Cavalleri}},\ }\bibfield  {title} {\bibinfo {title}
  {Amplification of superconducting fluctuations in driven
  {YBa$_{2}$Cu$_{3}$O$_{6+x}$}},\ }\href
  {https://doi.org/10.1103/PhysRevX.12.031008} {\bibfield  {journal} {\bibinfo
  {journal} {Phys. Rev. X}\ }\textbf {\bibinfo {volume} {12}},\ \bibinfo
  {pages} {031008} (\bibinfo {year} {2022})}\BibitemShut {NoStop}%
\bibitem [{\citenamefont {Michael}\ \emph {et~al.}(2020)\citenamefont
  {Michael}, \citenamefont {von Hoegen}, \citenamefont {Fechner}, \citenamefont
  {F\"orst}, \citenamefont {Cavalleri},\ and\ \citenamefont
  {Demler}}]{Michael2020}%
  \BibitemOpen
  \bibfield  {author} {\bibinfo {author} {\bibfnamefont {M.~H.}\ \bibnamefont
  {Michael}}, \bibinfo {author} {\bibfnamefont {A.}~\bibnamefont {von Hoegen}},
  \bibinfo {author} {\bibfnamefont {M.}~\bibnamefont {Fechner}}, \bibinfo
  {author} {\bibfnamefont {M.}~\bibnamefont {F\"orst}}, \bibinfo {author}
  {\bibfnamefont {A.}~\bibnamefont {Cavalleri}},\ and\ \bibinfo {author}
  {\bibfnamefont {E.}~\bibnamefont {Demler}},\ }\bibfield  {title} {\bibinfo
  {title} {Parametric resonance of {Josephson} plasma waves: A theory for
  optically amplified interlayer superconductivity in
  {${\mathrm{YBa}}_{2}{\mathrm{Cu}}_{3}{\mathrm{O}}_{6+x}$}},\ }\href
  {https://doi.org/10.1103/PhysRevB.102.174505} {\bibfield  {journal} {\bibinfo
   {journal} {Phys. Rev. B}\ }\textbf {\bibinfo {volume} {102}},\ \bibinfo
  {pages} {174505} (\bibinfo {year} {2020})}\BibitemShut {NoStop}%
\bibitem [{\citenamefont {Homann}\ \emph {et~al.}(2020)\citenamefont {Homann},
  \citenamefont {Cosme},\ and\ \citenamefont {Mathey}}]{Homann2020}%
  \BibitemOpen
  \bibfield  {author} {\bibinfo {author} {\bibfnamefont {G.}~\bibnamefont
  {Homann}}, \bibinfo {author} {\bibfnamefont {J.~G.}\ \bibnamefont {Cosme}},\
  and\ \bibinfo {author} {\bibfnamefont {L.}~\bibnamefont {Mathey}},\
  }\bibfield  {title} {\bibinfo {title} {Higgs time crystal in a high-${T}_{c}$
  superconductor},\ }\href {https://doi.org/10.1103/PhysRevResearch.2.043214}
  {\bibfield  {journal} {\bibinfo  {journal} {Phys. Rev. Research}\ }\textbf
  {\bibinfo {volume} {2}},\ \bibinfo {pages} {043214} (\bibinfo {year}
  {2020})}\BibitemShut {NoStop}%
\bibitem [{\citenamefont {Homann}\ \emph {et~al.}(2021)\citenamefont {Homann},
  \citenamefont {Cosme}, \citenamefont {Okamoto},\ and\ \citenamefont
  {Mathey}}]{Homann2021}%
  \BibitemOpen
  \bibfield  {author} {\bibinfo {author} {\bibfnamefont {G.}~\bibnamefont
  {Homann}}, \bibinfo {author} {\bibfnamefont {J.~G.}\ \bibnamefont {Cosme}},
  \bibinfo {author} {\bibfnamefont {J.}~\bibnamefont {Okamoto}},\ and\ \bibinfo
  {author} {\bibfnamefont {L.}~\bibnamefont {Mathey}},\ }\bibfield  {title}
  {\bibinfo {title} {Higgs mode mediated enhancement of interlayer transport in
  high-${T}_{c}$ cuprate superconductors},\ }\href
  {https://doi.org/10.1103/PhysRevB.103.224503} {\bibfield  {journal} {\bibinfo
   {journal} {Phys. Rev. B}\ }\textbf {\bibinfo {volume} {103}},\ \bibinfo
  {pages} {224503} (\bibinfo {year} {2021})}\BibitemShut {NoStop}%
\bibitem [{\citenamefont {Homann}\ \emph {et~al.}(2022)\citenamefont {Homann},
  \citenamefont {Cosme},\ and\ \citenamefont {Mathey}}]{Homann2022}%
  \BibitemOpen
  \bibfield  {author} {\bibinfo {author} {\bibfnamefont {G.}~\bibnamefont
  {Homann}}, \bibinfo {author} {\bibfnamefont {J.~G.}\ \bibnamefont {Cosme}},\
  and\ \bibinfo {author} {\bibfnamefont {L.}~\bibnamefont {Mathey}},\
  }\bibfield  {title} {\bibinfo {title} {Parametric control of {Meissner}
  screening in light-driven superconductors},\ }\href
  {https://doi.org/10.1088/1367-2630/ac9b83} {\bibfield  {journal} {\bibinfo
  {journal} {New J. Phys.}\ }\textbf {\bibinfo {volume} {24}},\ \bibinfo
  {pages} {113007} (\bibinfo {year} {2022})}\BibitemShut {NoStop}%
\bibitem [{\citenamefont {{Glazman}}\ and\ \citenamefont
  {{Koshelev}}(1990)}]{Glazman1990}%
  \BibitemOpen
  \bibfield  {author} {\bibinfo {author} {\bibfnamefont {L.~I.}\ \bibnamefont
  {{Glazman}}}\ and\ \bibinfo {author} {\bibfnamefont {A.~E.}\ \bibnamefont
  {{Koshelev}}},\ }\bibfield  {title} {\bibinfo {title} {Critical behavior of
  layered superconductors},\ }\href@noop {} {\bibfield  {journal} {\bibinfo
  {journal} {Sov. Phys. JETP}\ }\textbf {\bibinfo {volume} {70}},\ \bibinfo
  {pages} {774} (\bibinfo {year} {1990})}\BibitemShut {NoStop}%
\bibitem [{\citenamefont {Janke}\ and\ \citenamefont
  {Matsui}(1990)}]{Janke1990}%
  \BibitemOpen
  \bibfield  {author} {\bibinfo {author} {\bibfnamefont {W.}~\bibnamefont
  {Janke}}\ and\ \bibinfo {author} {\bibfnamefont {T.}~\bibnamefont {Matsui}},\
  }\bibfield  {title} {\bibinfo {title} {Crossover in the {$XY$} model from
  three to two dimensions},\ }\href {https://doi.org/10.1103/PhysRevB.42.10673}
  {\bibfield  {journal} {\bibinfo  {journal} {Phys. Rev. B}\ }\textbf {\bibinfo
  {volume} {42}},\ \bibinfo {pages} {10673} (\bibinfo {year}
  {1990})}\BibitemShut {NoStop}%
\bibitem [{\citenamefont {Minnhagen}\ and\ \citenamefont
  {Olsson}(1991{\natexlab{a}})}]{Minnhagen1991a}%
  \BibitemOpen
  \bibfield  {author} {\bibinfo {author} {\bibfnamefont {P.}~\bibnamefont
  {Minnhagen}}\ and\ \bibinfo {author} {\bibfnamefont {P.}~\bibnamefont
  {Olsson}},\ }\bibfield  {title} {\bibinfo {title} {Crossover to effectively
  two-dimensional vortices for high-{${\mathit{T}}_{\mathit{c}}$}
  superconductors},\ }\href {https://doi.org/10.1103/PhysRevLett.67.1039}
  {\bibfield  {journal} {\bibinfo  {journal} {Phys. Rev. Lett.}\ }\textbf
  {\bibinfo {volume} {67}},\ \bibinfo {pages} {1039} (\bibinfo {year}
  {1991}{\natexlab{a}})}\BibitemShut {NoStop}%
\bibitem [{\citenamefont {Minnhagen}\ and\ \citenamefont
  {Olsson}(1991{\natexlab{b}})}]{Minnhagen1991b}%
  \BibitemOpen
  \bibfield  {author} {\bibinfo {author} {\bibfnamefont {P.}~\bibnamefont
  {Minnhagen}}\ and\ \bibinfo {author} {\bibfnamefont {P.}~\bibnamefont
  {Olsson}},\ }\bibfield  {title} {\bibinfo {title} {Monte {Carlo} calculation
  of the vortex interaction for high-{${\mathit{T}}_{\mathit{c}}$}
  superconductors},\ }\href {https://doi.org/10.1103/PhysRevB.44.4503}
  {\bibfield  {journal} {\bibinfo  {journal} {Phys. Rev. B}\ }\textbf {\bibinfo
  {volume} {44}},\ \bibinfo {pages} {4503} (\bibinfo {year}
  {1991}{\natexlab{b}})}\BibitemShut {NoStop}%
\bibitem [{\citenamefont {Fisher}\ \emph {et~al.}(1991)\citenamefont {Fisher},
  \citenamefont {Fisher},\ and\ \citenamefont {Huse}}]{Fisher1991}%
  \BibitemOpen
  \bibfield  {author} {\bibinfo {author} {\bibfnamefont {D.~S.}\ \bibnamefont
  {Fisher}}, \bibinfo {author} {\bibfnamefont {M.~P.~A.}\ \bibnamefont
  {Fisher}},\ and\ \bibinfo {author} {\bibfnamefont {D.~A.}\ \bibnamefont
  {Huse}},\ }\bibfield  {title} {\bibinfo {title} {Thermal fluctuations,
  quenched disorder, phase transitions, and transport in type-{II}
  superconductors},\ }\href {https://doi.org/10.1103/PhysRevB.43.130}
  {\bibfield  {journal} {\bibinfo  {journal} {Phys. Rev. B}\ }\textbf {\bibinfo
  {volume} {43}},\ \bibinfo {pages} {130} (\bibinfo {year} {1991})}\BibitemShut
  {NoStop}%
\bibitem [{\citenamefont {Chattopadhyay}\ and\ \citenamefont
  {Shenoy}(1994)}]{Chattopadhyay1994}%
  \BibitemOpen
  \bibfield  {author} {\bibinfo {author} {\bibfnamefont {B.}~\bibnamefont
  {Chattopadhyay}}\ and\ \bibinfo {author} {\bibfnamefont {S.~R.}\ \bibnamefont
  {Shenoy}},\ }\bibfield  {title} {\bibinfo {title} {{Kosterlitz-Thouless}
  signatures from {3D} vortex loops in layered superconductors},\ }\href
  {https://doi.org/10.1103/PhysRevLett.72.400} {\bibfield  {journal} {\bibinfo
  {journal} {Phys. Rev. Lett.}\ }\textbf {\bibinfo {volume} {72}},\ \bibinfo
  {pages} {400} (\bibinfo {year} {1994})}\BibitemShut {NoStop}%
\bibitem [{\citenamefont {{Mathey}}\ \emph {et~al.}(2008)\citenamefont
  {{Mathey}}, \citenamefont {{Polkovnikov}},\ and\ \citenamefont {{Castro
  Neto}}}]{Mathey2008}%
  \BibitemOpen
  \bibfield  {author} {\bibinfo {author} {\bibfnamefont {L.}~\bibnamefont
  {{Mathey}}}, \bibinfo {author} {\bibfnamefont {A.}~\bibnamefont
  {{Polkovnikov}}},\ and\ \bibinfo {author} {\bibfnamefont {A.~H.}\
  \bibnamefont {{Castro Neto}}},\ }\bibfield  {title} {\bibinfo {title}
  {Phase-locking transition of coupled low-dimensional superfluids},\ }\href
  {https://doi.org/10.1209/0295-5075/81/10008} {\bibfield  {journal} {\bibinfo
  {journal} {EPL}\ }\textbf {\bibinfo {volume} {81}},\ \bibinfo {pages} {10008}
  (\bibinfo {year} {2008})}\BibitemShut {NoStop}%
\bibitem [{\citenamefont {Hu}\ \emph {et~al.}(2009)\citenamefont {Hu},
  \citenamefont {Mathey}, \citenamefont {Danshita}, \citenamefont {Tiesinga},
  \citenamefont {Williams},\ and\ \citenamefont {Clark}}]{Hu2009}%
  \BibitemOpen
  \bibfield  {author} {\bibinfo {author} {\bibfnamefont {A.}~\bibnamefont
  {Hu}}, \bibinfo {author} {\bibfnamefont {L.}~\bibnamefont {Mathey}}, \bibinfo
  {author} {\bibfnamefont {I.}~\bibnamefont {Danshita}}, \bibinfo {author}
  {\bibfnamefont {E.}~\bibnamefont {Tiesinga}}, \bibinfo {author}
  {\bibfnamefont {C.~J.}\ \bibnamefont {Williams}},\ and\ \bibinfo {author}
  {\bibfnamefont {C.~W.}\ \bibnamefont {Clark}},\ }\bibfield  {title} {\bibinfo
  {title} {Counterflow and paired superfluidity in one-dimensional {Bose}
  mixtures in optical lattices},\ }\href
  {https://doi.org/10.1103/PhysRevA.80.023619} {\bibfield  {journal} {\bibinfo
  {journal} {Phys. Rev. A}\ }\textbf {\bibinfo {volume} {80}},\ \bibinfo
  {pages} {023619} (\bibinfo {year} {2009})}\BibitemShut {NoStop}%
\bibitem [{\citenamefont {Hu}\ \emph {et~al.}(2011)\citenamefont {Hu},
  \citenamefont {Mathey}, \citenamefont {Tiesinga}, \citenamefont {Danshita},
  \citenamefont {Williams},\ and\ \citenamefont {Clark}}]{Hu2011}%
  \BibitemOpen
  \bibfield  {author} {\bibinfo {author} {\bibfnamefont {A.}~\bibnamefont
  {Hu}}, \bibinfo {author} {\bibfnamefont {L.}~\bibnamefont {Mathey}}, \bibinfo
  {author} {\bibfnamefont {E.}~\bibnamefont {Tiesinga}}, \bibinfo {author}
  {\bibfnamefont {I.}~\bibnamefont {Danshita}}, \bibinfo {author}
  {\bibfnamefont {C.~J.}\ \bibnamefont {Williams}},\ and\ \bibinfo {author}
  {\bibfnamefont {C.~W.}\ \bibnamefont {Clark}},\ }\bibfield  {title} {\bibinfo
  {title} {Detecting paired and counterflow superfluidity via dipole
  oscillations},\ }\href {https://doi.org/10.1103/PhysRevA.84.041609}
  {\bibfield  {journal} {\bibinfo  {journal} {Phys. Rev. A}\ }\textbf {\bibinfo
  {volume} {84}},\ \bibinfo {pages} {041609} (\bibinfo {year}
  {2011})}\BibitemShut {NoStop}%
\bibitem [{\citenamefont {Eisenstein}\ \emph {et~al.}(1992)\citenamefont
  {Eisenstein}, \citenamefont {Boebinger}, \citenamefont {Pfeiffer},
  \citenamefont {West},\ and\ \citenamefont {He}}]{Eisenstein1992}%
  \BibitemOpen
  \bibfield  {author} {\bibinfo {author} {\bibfnamefont {J.~P.}\ \bibnamefont
  {Eisenstein}}, \bibinfo {author} {\bibfnamefont {G.~S.}\ \bibnamefont
  {Boebinger}}, \bibinfo {author} {\bibfnamefont {L.~N.}\ \bibnamefont
  {Pfeiffer}}, \bibinfo {author} {\bibfnamefont {K.~W.}\ \bibnamefont {West}},\
  and\ \bibinfo {author} {\bibfnamefont {S.}~\bibnamefont {He}},\ }\bibfield
  {title} {\bibinfo {title} {New fractional quantum {Hall} state in
  double-layer two-dimensional electron systems},\ }\href
  {https://doi.org/10.1103/PhysRevLett.68.1383} {\bibfield  {journal} {\bibinfo
   {journal} {Phys. Rev. Lett.}\ }\textbf {\bibinfo {volume} {68}},\ \bibinfo
  {pages} {1383} (\bibinfo {year} {1992})}\BibitemShut {NoStop}%
\bibitem [{\citenamefont {Tutuc}\ \emph {et~al.}(2004)\citenamefont {Tutuc},
  \citenamefont {Shayegan},\ and\ \citenamefont {Huse}}]{Tutuc2004}%
  \BibitemOpen
  \bibfield  {author} {\bibinfo {author} {\bibfnamefont {E.}~\bibnamefont
  {Tutuc}}, \bibinfo {author} {\bibfnamefont {M.}~\bibnamefont {Shayegan}},\
  and\ \bibinfo {author} {\bibfnamefont {D.~A.}\ \bibnamefont {Huse}},\
  }\bibfield  {title} {\bibinfo {title} {Counterflow measurements in strongly
  correlated {GaAs} hole bilayers: Evidence for electron-hole pairing},\ }\href
  {https://doi.org/10.1103/PhysRevLett.93.036802} {\bibfield  {journal}
  {\bibinfo  {journal} {Phys. Rev. Lett.}\ }\textbf {\bibinfo {volume} {93}},\
  \bibinfo {pages} {036802} (\bibinfo {year} {2004})}\BibitemShut {NoStop}%
\bibitem [{\citenamefont {{Eisenstein}}\ and\ \citenamefont
  {{MacDonald}}(2004)}]{Eisenstein2004}%
  \BibitemOpen
  \bibfield  {author} {\bibinfo {author} {\bibfnamefont {J.~P.}\ \bibnamefont
  {{Eisenstein}}}\ and\ \bibinfo {author} {\bibfnamefont {A.~H.}\ \bibnamefont
  {{MacDonald}}},\ }\bibfield  {title} {\bibinfo {title} {{Bose-Einstein}
  condensation of excitons in bilayer electron systems},\ }\href
  {https://doi.org/10.1038/nature03081} {\bibfield  {journal} {\bibinfo
  {journal} {Nature}\ }\textbf {\bibinfo {volume} {432}},\ \bibinfo {pages}
  {691} (\bibinfo {year} {2004})}\BibitemShut {NoStop}%
\bibitem [{\citenamefont {{Su}}\ and\ \citenamefont
  {{MacDonald}}(2008)}]{Su2008}%
  \BibitemOpen
  \bibfield  {author} {\bibinfo {author} {\bibfnamefont {J.-J.}\ \bibnamefont
  {{Su}}}\ and\ \bibinfo {author} {\bibfnamefont {A.~H.}\ \bibnamefont
  {{MacDonald}}},\ }\bibfield  {title} {\bibinfo {title} {How to make a bilayer
  exciton condensate flow},\ }\href {https://doi.org/10.1038/nphys1055}
  {\bibfield  {journal} {\bibinfo  {journal} {Nat. Phys.}\ }\textbf {\bibinfo
  {volume} {4}},\ \bibinfo {pages} {799} (\bibinfo {year} {2008})}\BibitemShut
  {NoStop}%
\bibitem [{\citenamefont {Kobayashi}\ \emph {et~al.}(2019)\citenamefont
  {Kobayashi}, \citenamefont {Yui},\ and\ \citenamefont
  {Tsubota}}]{Kobayashi19}%
  \BibitemOpen
  \bibfield  {author} {\bibinfo {author} {\bibfnamefont {H.}~\bibnamefont
  {Kobayashi}}, \bibinfo {author} {\bibfnamefont {S.}~\bibnamefont {Yui}},\
  and\ \bibinfo {author} {\bibfnamefont {M.}~\bibnamefont {Tsubota}},\
  }\bibfield  {title} {\bibinfo {title} {Numerical study on entrance length in
  thermal counterflow of superfluid {\$}{\$}\^{}4{\$}{\$}he},\ }\href
  {https://doi.org/10.1007/s10909-019-02169-8} {\bibfield  {journal} {\bibinfo
  {journal} {Journal of Low Temperature Physics}\ }\textbf {\bibinfo {volume}
  {196}},\ \bibinfo {pages} {35} (\bibinfo {year} {2019})}\BibitemShut
  {NoStop}%
\bibitem [{\citenamefont {Ginzburg}\ and\ \citenamefont
  {Landau}(1950)}]{Ginzburg1950}%
  \BibitemOpen
  \bibfield  {author} {\bibinfo {author} {\bibfnamefont {V.~L.}\ \bibnamefont
  {Ginzburg}}\ and\ \bibinfo {author} {\bibfnamefont {L.~D.}\ \bibnamefont
  {Landau}},\ }\bibfield  {title} {\bibinfo {title} {On the theory of
  superconductivity},\ }\href@noop {} {\bibfield  {journal} {\bibinfo
  {journal} {Zh. Eksp. Teor. Fiz.}\ }\textbf {\bibinfo {volume} {20}},\
  \bibinfo {pages} {1064} (\bibinfo {year} {1950})}\BibitemShut {NoStop}%
\bibitem [{Sup()}]{Supp}%
  \BibitemOpen
  \href@noop {} {}\bibinfo {note} {{See Supplemental Material for details on
  the lattice gauge model, model parameters, the temperature dependence of the
  in-plane tunneling and the superconducting order parameter, information on
  the plasma resonances, vortex correlations, and details on the conductivity
  measurements.}}\BibitemShut {Stop}%
\bibitem [{\citenamefont {Halperin}\ and\ \citenamefont
  {Nelson}(1979)}]{Halperin1979}%
  \BibitemOpen
  \bibfield  {author} {\bibinfo {author} {\bibfnamefont {B.}~\bibnamefont
  {Halperin}}\ and\ \bibinfo {author} {\bibfnamefont {D.}~\bibnamefont
  {Nelson}},\ }\bibfield  {title} {\bibinfo {title} {Resistive transition in
  superconducting films},\ }\href {https://doi.org/10.1007/BF00116988}
  {\bibfield  {journal} {\bibinfo  {journal} {J. Low Temp. Phys.}\ }\textbf
  {\bibinfo {volume} {36}},\ \bibinfo {pages} {599} (\bibinfo {year}
  {1979})}\BibitemShut {NoStop}%
\bibitem [{\citenamefont {Minnhagen}(1987)}]{Minnhagen1987}%
  \BibitemOpen
  \bibfield  {author} {\bibinfo {author} {\bibfnamefont {P.}~\bibnamefont
  {Minnhagen}},\ }\bibfield  {title} {\bibinfo {title} {The two-dimensional
  {Coulomb} gas, vortex unbinding, and superfluid-superconducting films},\
  }\href {https://doi.org/10.1103/RevModPhys.59.1001} {\bibfield  {journal}
  {\bibinfo  {journal} {Rev. Mod. Phys.}\ }\textbf {\bibinfo {volume} {59}},\
  \bibinfo {pages} {1001} (\bibinfo {year} {1987})}\BibitemShut {NoStop}%
\bibitem [{\citenamefont {Girvin}\ and\ \citenamefont
  {MacDonald}(1996)}]{Girvin96}%
  \BibitemOpen
  \bibfield  {author} {\bibinfo {author} {\bibfnamefont {S.~M.}\ \bibnamefont
  {Girvin}}\ and\ \bibinfo {author} {\bibfnamefont {A.~H.}\ \bibnamefont
  {MacDonald}},\ }\bibinfo {title} {Multicomponent quantum hall systems: The
  sum of their parts and more},\ in\ \href
  {https://doi.org/https://doi.org/10.1002/9783527617258.ch5} {\emph {\bibinfo
  {booktitle} {Perspectives in Quantum Hall Effects}}}\ (\bibinfo  {publisher}
  {John Wiley \& Sons, Ltd},\ \bibinfo {year} {1996})\ Chap.~\bibinfo {chapter}
  {5}, pp.\ \bibinfo {pages} {161--224}\BibitemShut {NoStop}%
\bibitem [{\citenamefont {Yu}\ \emph {et~al.}(2016)\citenamefont {Yu},
  \citenamefont {Hirschberger}, \citenamefont {Loew}, \citenamefont {Li},
  \citenamefont {Lawson}, \citenamefont {Asaba}, \citenamefont {Kemper},
  \citenamefont {Liang}, \citenamefont {Porras}, \citenamefont {Boebinger},
  \citenamefont {Singleton}, \citenamefont {Keimer}, \citenamefont {Li},\ and\
  \citenamefont {Ong}}]{Fan16}%
  \BibitemOpen
  \bibfield  {author} {\bibinfo {author} {\bibfnamefont {F.}~\bibnamefont
  {Yu}}, \bibinfo {author} {\bibfnamefont {M.}~\bibnamefont {Hirschberger}},
  \bibinfo {author} {\bibfnamefont {T.}~\bibnamefont {Loew}}, \bibinfo {author}
  {\bibfnamefont {G.}~\bibnamefont {Li}}, \bibinfo {author} {\bibfnamefont
  {B.~J.}\ \bibnamefont {Lawson}}, \bibinfo {author} {\bibfnamefont
  {T.}~\bibnamefont {Asaba}}, \bibinfo {author} {\bibfnamefont {J.~B.}\
  \bibnamefont {Kemper}}, \bibinfo {author} {\bibfnamefont {T.}~\bibnamefont
  {Liang}}, \bibinfo {author} {\bibfnamefont {J.}~\bibnamefont {Porras}},
  \bibinfo {author} {\bibfnamefont {G.~S.}\ \bibnamefont {Boebinger}}, \bibinfo
  {author} {\bibfnamefont {J.}~\bibnamefont {Singleton}}, \bibinfo {author}
  {\bibfnamefont {B.}~\bibnamefont {Keimer}}, \bibinfo {author} {\bibfnamefont
  {L.}~\bibnamefont {Li}},\ and\ \bibinfo {author} {\bibfnamefont {N.~P.}\
  \bibnamefont {Ong}},\ }\bibfield  {title} {\bibinfo {title} {Magnetic phase
  diagram of underdoped {YBa$_2$Cu$_3$O$_y$} inferred from torque magnetization
  and thermal conductivity},\ }\href {https://doi.org/10.1073/pnas.1612591113}
  {\bibfield  {journal} {\bibinfo  {journal} {Proc. Natl. Acad. Sci.}\ }\textbf
  {\bibinfo {volume} {113}},\ \bibinfo {pages} {12667} (\bibinfo {year}
  {2016})}\BibitemShut {NoStop}%
\end{thebibliography}%


\providecommand{\noopsort}[1]{}\providecommand{\singleletter}[1]{#1}%
\begin{thebibliography}{11}%
\makeatletter
\providecommand \@ifxundefined [1]{%
 \@ifx{#1\undefined}
}%
\providecommand \@ifnum [1]{%
 \ifnum #1\expandafter \@firstoftwo
 \else \expandafter \@secondoftwo
 \fi
}%
\providecommand \@ifx [1]{%
 \ifx #1\expandafter \@firstoftwo
 \else \expandafter \@secondoftwo
 \fi
}%
\providecommand \natexlab [1]{#1}%
\providecommand \enquote  [1]{``#1''}%
\providecommand \bibnamefont  [1]{#1}%
\providecommand \bibfnamefont [1]{#1}%
\providecommand \citenamefont [1]{#1}%
\providecommand \href@noop [0]{\@secondoftwo}%
\providecommand \href [0]{\begingroup \@sanitize@url \@href}%
\providecommand \@href[1]{\@@startlink{#1}\@@href}%
\providecommand \@@href[1]{\endgroup#1\@@endlink}%
\providecommand \@sanitize@url [0]{\catcode `\\12\catcode `\$12\catcode
  `\&12\catcode `\#12\catcode `\^12\catcode `\_12\catcode `\%12\relax}%
\providecommand \@@startlink[1]{}%
\providecommand \@@endlink[0]{}%
\providecommand \url  [0]{\begingroup\@sanitize@url \@url }%
\providecommand \@url [1]{\endgroup\@href {#1}{\urlprefix }}%
\providecommand \urlprefix  [0]{URL }%
\providecommand \Eprint [0]{\href }%
\providecommand \doibase [0]{https://doi.org/}%
\providecommand \selectlanguage [0]{\@gobble}%
\providecommand \bibinfo  [0]{\@secondoftwo}%
\providecommand \bibfield  [0]{\@secondoftwo}%
\providecommand \translation [1]{[#1]}%
\providecommand \BibitemOpen [0]{}%
\providecommand \bibitemStop [0]{}%
\providecommand \bibitemNoStop [0]{.\EOS\space}%
\providecommand \EOS [0]{\spacefactor3000\relax}%
\providecommand \BibitemShut  [1]{\csname bibitem#1\endcsname}%
\let\auto@bib@innerbib\@empty
\bibitem [{\citenamefont {Homann}\ \emph {et~al.}(2020)\citenamefont {Homann},
  \citenamefont {Cosme},\ and\ \citenamefont {Mathey}}]{Homann2020}%
  \BibitemOpen
  \bibfield  {author} {\bibinfo {author} {\bibfnamefont {G.}~\bibnamefont
  {Homann}}, \bibinfo {author} {\bibfnamefont {J.~G.}\ \bibnamefont {Cosme}},\
  and\ \bibinfo {author} {\bibfnamefont {L.}~\bibnamefont {Mathey}},\
  }\bibfield  {title} {\bibinfo {title} {Higgs time crystal in a high-${T}_{c}$
  superconductor},\ }\href {https://doi.org/10.1103/PhysRevResearch.2.043214}
  {\bibfield  {journal} {\bibinfo  {journal} {Phys. Rev. Research}\ }\textbf
  {\bibinfo {volume} {2}},\ \bibinfo {pages} {043214} (\bibinfo {year}
  {2020})}\BibitemShut {NoStop}%
\bibitem [{\citenamefont {Homann}\ \emph {et~al.}(2021)\citenamefont {Homann},
  \citenamefont {Cosme}, \citenamefont {Okamoto},\ and\ \citenamefont
  {Mathey}}]{Homann2021}%
  \BibitemOpen
  \bibfield  {author} {\bibinfo {author} {\bibfnamefont {G.}~\bibnamefont
  {Homann}}, \bibinfo {author} {\bibfnamefont {J.~G.}\ \bibnamefont {Cosme}},
  \bibinfo {author} {\bibfnamefont {J.}~\bibnamefont {Okamoto}},\ and\ \bibinfo
  {author} {\bibfnamefont {L.}~\bibnamefont {Mathey}},\ }\bibfield  {title}
  {\bibinfo {title} {Higgs mode mediated enhancement of interlayer transport in
  high-${T}_{c}$ cuprate superconductors},\ }\href
  {https://doi.org/10.1103/PhysRevB.103.224503} {\bibfield  {journal} {\bibinfo
   {journal} {Phys. Rev. B}\ }\textbf {\bibinfo {volume} {103}},\ \bibinfo
  {pages} {224503} (\bibinfo {year} {2021})}\BibitemShut {NoStop}%
\bibitem [{\citenamefont {Homann}\ \emph {et~al.}(2022)\citenamefont {Homann},
  \citenamefont {Cosme},\ and\ \citenamefont {Mathey}}]{Homann2022}%
  \BibitemOpen
  \bibfield  {author} {\bibinfo {author} {\bibfnamefont {G.}~\bibnamefont
  {Homann}}, \bibinfo {author} {\bibfnamefont {J.~G.}\ \bibnamefont {Cosme}},\
  and\ \bibinfo {author} {\bibfnamefont {L.}~\bibnamefont {Mathey}},\
  }\bibfield  {title} {\bibinfo {title} {Parametric control of {Meissner}
  screening in light-driven superconductors},\ }\href
  {https://doi.org/10.1088/1367-2630/ac9b83} {\bibfield  {journal} {\bibinfo
  {journal} {New J. Phys.}\ }\textbf {\bibinfo {volume} {24}},\ \bibinfo
  {pages} {113007} (\bibinfo {year} {2022})}\BibitemShut {NoStop}%
\bibitem [{\citenamefont {Machida}\ \emph {et~al.}(1999)\citenamefont
  {Machida}, \citenamefont {Koyama},\ and\ \citenamefont
  {Tachiki}}]{Machida1999}%
  \BibitemOpen
  \bibfield  {author} {\bibinfo {author} {\bibfnamefont {M.}~\bibnamefont
  {Machida}}, \bibinfo {author} {\bibfnamefont {T.}~\bibnamefont {Koyama}},\
  and\ \bibinfo {author} {\bibfnamefont {M.}~\bibnamefont {Tachiki}},\
  }\bibfield  {title} {\bibinfo {title} {Dynamical breaking of charge
  neutrality in intrinsic {Josephson} junctions: Common origin for microwave
  resonant absorptions and multiple-branch structures in the
  $\mathit{I}\ensuremath{-}\mathit{V}$ characteristics},\ }\href
  {https://doi.org/10.1103/PhysRevLett.83.4618} {\bibfield  {journal} {\bibinfo
   {journal} {Phys. Rev. Lett.}\ }\textbf {\bibinfo {volume} {83}},\ \bibinfo
  {pages} {4618} (\bibinfo {year} {1999})}\BibitemShut {NoStop}%
\bibitem [{\citenamefont {van~der Marel}\ and\ \citenamefont
  {Tsvetkov}(2001)}]{VanDerMarel2001}%
  \BibitemOpen
  \bibfield  {author} {\bibinfo {author} {\bibfnamefont {D.}~\bibnamefont
  {van~der Marel}}\ and\ \bibinfo {author} {\bibfnamefont {A.~A.}\ \bibnamefont
  {Tsvetkov}},\ }\bibfield  {title} {\bibinfo {title} {Transverse-optical
  {Josephson} plasmons: Equations of motion},\ }\href
  {https://doi.org/10.1103/PhysRevB.64.024530} {\bibfield  {journal} {\bibinfo
  {journal} {Phys. Rev. B}\ }\textbf {\bibinfo {volume} {64}},\ \bibinfo
  {pages} {024530} (\bibinfo {year} {2001})}\BibitemShut {NoStop}%
\bibitem [{\citenamefont {{Koyama}}(2002)}]{Koyama2002}%
  \BibitemOpen
  \bibfield  {author} {\bibinfo {author} {\bibfnamefont {T.}~\bibnamefont
  {{Koyama}}},\ }\bibfield  {title} {\bibinfo {title} {Josephson plasma
  resonances and optical properties in high-${T}_{c}$ superconductors with
  alternating junction parameters},\ }\href
  {https://doi.org/10.1143/JPSJ.71.2986} {\bibfield  {journal} {\bibinfo
  {journal} {J. Phys. Soc. Jpn.}\ }\textbf {\bibinfo {volume} {71}},\ \bibinfo
  {pages} {2986} (\bibinfo {year} {2002})}\BibitemShut {NoStop}%
\bibitem [{\citenamefont {Koyama}\ and\ \citenamefont
  {Tachiki}(1996)}]{Koyama1996}%
  \BibitemOpen
  \bibfield  {author} {\bibinfo {author} {\bibfnamefont {T.}~\bibnamefont
  {Koyama}}\ and\ \bibinfo {author} {\bibfnamefont {M.}~\bibnamefont
  {Tachiki}},\ }\bibfield  {title} {\bibinfo {title} {${I}$-${V}$
  characteristics of {Josephson}-coupled layered superconductors with
  longitudinal plasma excitations},\ }\href
  {https://doi.org/10.1103/PhysRevB.54.16183} {\bibfield  {journal} {\bibinfo
  {journal} {Phys. Rev. B}\ }\textbf {\bibinfo {volume} {54}},\ \bibinfo
  {pages} {16183} (\bibinfo {year} {1996})}\BibitemShut {NoStop}%
\bibitem [{\citenamefont {Machida}\ and\ \citenamefont
  {Koyama}(2004)}]{Machida2004}%
  \BibitemOpen
  \bibfield  {author} {\bibinfo {author} {\bibfnamefont {M.}~\bibnamefont
  {Machida}}\ and\ \bibinfo {author} {\bibfnamefont {T.}~\bibnamefont
  {Koyama}},\ }\bibfield  {title} {\bibinfo {title} {Localized rotating-modes
  in capacitively coupled intrinsic {Josephson} junctions: Systematic study of
  branching structure and collective dynamical instability},\ }\href
  {https://doi.org/10.1103/PhysRevB.70.024523} {\bibfield  {journal} {\bibinfo
  {journal} {Phys. Rev. B}\ }\textbf {\bibinfo {volume} {70}},\ \bibinfo
  {pages} {024523} (\bibinfo {year} {2004})}\BibitemShut {NoStop}%
\bibitem [{\citenamefont {Shibata}\ and\ \citenamefont
  {Yamada}(1998)}]{Shibata1998}%
  \BibitemOpen
  \bibfield  {author} {\bibinfo {author} {\bibfnamefont {H.}~\bibnamefont
  {Shibata}}\ and\ \bibinfo {author} {\bibfnamefont {T.}~\bibnamefont
  {Yamada}},\ }\bibfield  {title} {\bibinfo {title} {Double {Josephson} plasma
  resonance in {$T^{*}$} phase
  {${\mathrm{SmLa}}_{1\ensuremath{-}\mathit{x}}{\mathrm{Sr}}_{\mathit{x}}{\mathrm{CuO}}_{4\ensuremath{-}\mathit{\ensuremath{\delta}}}$}},\
  }\href {https://doi.org/10.1103/PhysRevLett.81.3519} {\bibfield  {journal}
  {\bibinfo  {journal} {Phys. Rev. Lett.}\ }\textbf {\bibinfo {volume} {81}},\
  \bibinfo {pages} {3519} (\bibinfo {year} {1998})}\BibitemShut {NoStop}%
\bibitem [{\citenamefont {von Hoegen}\ \emph {et~al.}(2022)\citenamefont {von
  Hoegen}, \citenamefont {Fechner}, \citenamefont {F\"orst}, \citenamefont
  {Taherian}, \citenamefont {Rowe}, \citenamefont {Ribak}, \citenamefont
  {Porras}, \citenamefont {Keimer}, \citenamefont {Michael}, \citenamefont
  {Demler},\ and\ \citenamefont {Cavalleri}}]{VonHoegen2022}%
  \BibitemOpen
  \bibfield  {author} {\bibinfo {author} {\bibfnamefont {A.}~\bibnamefont {von
  Hoegen}}, \bibinfo {author} {\bibfnamefont {M.}~\bibnamefont {Fechner}},
  \bibinfo {author} {\bibfnamefont {M.}~\bibnamefont {F\"orst}}, \bibinfo
  {author} {\bibfnamefont {N.}~\bibnamefont {Taherian}}, \bibinfo {author}
  {\bibfnamefont {E.}~\bibnamefont {Rowe}}, \bibinfo {author} {\bibfnamefont
  {A.}~\bibnamefont {Ribak}}, \bibinfo {author} {\bibfnamefont
  {J.}~\bibnamefont {Porras}}, \bibinfo {author} {\bibfnamefont
  {B.}~\bibnamefont {Keimer}}, \bibinfo {author} {\bibfnamefont
  {M.}~\bibnamefont {Michael}}, \bibinfo {author} {\bibfnamefont
  {E.}~\bibnamefont {Demler}},\ and\ \bibinfo {author} {\bibfnamefont
  {A.}~\bibnamefont {Cavalleri}},\ }\bibfield  {title} {\bibinfo {title}
  {Amplification of superconducting fluctuations in driven
  {YBa$_{2}$Cu$_{3}$O$_{6+x}$}},\ }\href
  {https://doi.org/10.1103/PhysRevX.12.031008} {\bibfield  {journal} {\bibinfo
  {journal} {Phys. Rev. X}\ }\textbf {\bibinfo {volume} {12}},\ \bibinfo
  {pages} {031008} (\bibinfo {year} {2022})}\BibitemShut {NoStop}%
\bibitem [{\citenamefont {Schneider}\ and\ \citenamefont
  {Stoll}(1976)}]{Schneider1976}%
  \BibitemOpen
  \bibfield  {author} {\bibinfo {author} {\bibfnamefont {T.}~\bibnamefont
  {Schneider}}\ and\ \bibinfo {author} {\bibfnamefont {E.}~\bibnamefont
  {Stoll}},\ }\bibfield  {title} {\bibinfo {title} {Molecular-dynamics study of
  structural-phase transitions. {I. One}-component displacement models},\
  }\href {https://doi.org/10.1103/PhysRevB.13.1216} {\bibfield  {journal}
  {\bibinfo  {journal} {Phys. Rev. B}\ }\textbf {\bibinfo {volume} {13}},\
  \bibinfo {pages} {1216} (\bibinfo {year} {1976})}\BibitemShut {NoStop}%
\end{thebibliography}%

\end{document}


\title{Supplemental Material for\\Dissipationless counterflow above $T_c$ in bilayer superconductors}

\author{Guido Homann}
\affiliation{Zentrum f\"ur Optische Quantentechnologien and Institut f\"ur Quantenphysik, 
	Universit\"at Hamburg, 22761 Hamburg, Germany}

\author{Marios H. Michael}
\affiliation{Max Planck Institute for the Structure and Dynamics of Matter, Luruper Chausse 149, 22761 Hamburg, Germany}

\author{Jayson G. Cosme}
\affiliation{National Institute of Physics, University of the Philippines, Diliman, Quezon City 1101, Philippines}

\author{Ludwig Mathey}
\affiliation{Zentrum f\"ur Optische Quantentechnologien and Institut f\"ur Laserphysik, 
	Universit\"at Hamburg, 22761 Hamburg, Germany}
\affiliation{The Hamburg Centre for Ultrafast Imaging, Luruper Chaussee 149, 22761 Hamburg, Germany}

\maketitle
\tableofcontents

\clearpage
\section{Lagrangian and equations of motion}
Here, we present our semiclassical $U(1)$ lattice gauge theory \cite{Homann2020, Homann2021, Homann2022} in detail. The in-plane lattice constant $d_{x,\mathbf{r}}= d_{y,\mathbf{r}}= d_{ab}$ is introduced as a short-range cutoff below the in-plane coherence length. The interlayer distances are $d_{z,\mathbf{r}}= d_s$ for intrabilayer (strong) junctions and $d_{z,\mathbf{r}}= d_w$ for interbilayer (weak) junctions, reproducing the spacing of CuO$_2$ planes in the crystal.
The Lagrangian of the lattice gauge model consists of three terms,
\begin{equation} \label{eq:Lagrangian}
	\mathcal{L} = \mathcal{L}_{\mathrm{sc}} + \mathcal{L}_{\mathrm{em}} + \mathcal{L}_{\mathrm{kin}} .
\end{equation}
The first term is the $|\psi|^4$ model of the superconducting condensate in the absence of Cooper pair tunneling,
\begin{equation}\label{eq:LSC}
	\mathcal{L}_{\mathrm{sc}} = \sum_{\mathbf{r}} K \hbar^2 | \partial_{t} \psi_{\mathbf{r}} |^2 + \mu | \psi_{\mathbf{r}} |^2 - \frac{g}{2} | \psi_{\mathbf{r}} |^4 ,
\end{equation}
where the Ginzburg-Landau coefficients $\mu$ and $g$ are kept fixed throughout this work. The coefficient $K$ is related to the Thomas-Fermi screening length $\lambda_{\mathrm{TF}}$ \cite{Machida1999}, $K= \epsilon_0/8e^2 |\psi_0|^2 \lambda_{\mathrm{TF}}^2$.
We formulate the Lagrangian of the free electromagnetic field on an anisotropic lattice,
\begin{equation} \label{eq:LEM}
	\mathcal{L}_{\mathrm{em}} = \sum_{j,\mathbf{r}} \frac{\kappa_{j,\mathbf{r}} \epsilon_\infty \epsilon_0}{2} E_{j,\mathbf{r}}^2 - \frac{\kappa_{z,\mathbf{r}}}{\kappa_{j,\mathbf{r}} \beta_{j,\mathbf{r}}^2 \mu_0} \Bigl[1 - \cos\bigl(\beta_{j,\mathbf{r}} B_{j,\mathbf{r}} \bigr) \Bigr] .
\end{equation}
We employ the temporal gauge, where the electric field is given by the time derivative of the vector potential, $E_{j,\mathbf{r}}= -\partial_t A_{j,\mathbf{r}}$. The $j$~component of the electric field lies on the bond from site $\mathbf{r}$ to its nearest neighbor in the $j \in \{x,y,z\}$ direction. The magnetic field components $B_{j,\mathbf{r}}= \epsilon_{jkl}\delta_k A_{l,\mathbf{r}}$ are centered on the plaquettes of the lattice. We calculate the spatial derivatives according to $\delta_k A_{l,\mathbf{r}} = (A_{l,\mathbf{r} + \mathbf{u}_k}-A_{l,\mathbf{r}})/d_{l,\mathbf{r}}$, where $\mathbf{u}_k$ is the unit vector in the $k$~direction.
The background permittivity $\epsilon_\infty$ is due to bound charges. The other prefactors in Eq.~\eqref{eq:LEM} are linked to the anisotropic lattice geometry. Introducing $d_c = (d_s + d_w)/2$, we write $\kappa_{x,\mathbf{r}}= \kappa_{y,\mathbf{r}}= 1$ and $\kappa_{z,\mathbf{r}}= d_{z,\mathbf{r}}/d_c$, while $\beta_{x,\mathbf{r}}= \beta_{y,\mathbf{r}}= 2ed_{ab}d_{z,\mathbf{r}}/\hbar$ and $\beta_{z,\mathbf{r}}= 2ed_{ab}^2/\hbar$.
The kinetic part of the Lagrangian,
\begin{equation} \label{eq:LEK}
	\mathcal{L}_{\mathrm{kin}} = - \sum_{j,\mathbf{r}} t_{j,\mathbf{r}} |\psi_{\mathbf{r} + \mathbf{u}_j} - \psi_{\mathbf{r}} e^{i a_{j,\mathbf{r}}}|^2 ,
\end{equation}
accounts for nearest-neighbor tunneling of Cooper pairs.
The unitless vector potential $a_{j,\mathbf{r}}= -2e d_{j,\mathbf{r}} A_{j,\mathbf{r}}/\hbar$ couples to the phase of the order parameter, ensuring the local gauge-invariance of $\mathcal{L}_{\mathrm{kin}}$. This coupling reflects the Coulomb interaction between the Cooper pairs. The tunneling coefficients are $t_{x,\mathbf{r}}= t_{y,\mathbf{r}}= t_{ab}$ for in-plane junctions, $t_{z,\mathbf{r}}= t_s$ for intrabilayer junctions, and $t_{z,\mathbf{r}}= t_w$ for interbilayer junctions.

The equations of motion read
\begin{align}
	\partial_t^2 \psi_{\mathbf{r}} &= \frac{1}{K \hbar^2} \frac{\partial \mathcal{L}}{\partial \psi_{\mathbf{r}}^*} - \gamma_{\mathrm{sc}} \partial_t \psi_{\mathbf{r}} + \xi_{\mathbf{r}} , \\
	\partial_t^2 A_{j, \mathbf{r}} &= \frac{1}{\epsilon_\infty \epsilon_0} \frac{\partial \mathcal{L}}{\partial A_{j, \mathbf{r}}} - \gamma_{j,\mathbf{r}} \partial_t A_{j, \mathbf{r}} + \eta_{j, \mathbf{r}} ,
\end{align}
where $\gamma_{\mathrm{sc}}$ and $\gamma_{j,\mathbf{r}}$ are phenomenological damping constants of the superconducting order parameter and the vector potential, respectively. The damping constants of the vector potential are $\gamma_{x,\mathbf{r}}= \gamma_{y,\mathbf{r}}= \gamma_{ab}$ for in-plane junctions, $\gamma_{z,\mathbf{r}}= \gamma_s$ for intrabilayer junctions, and $\gamma_{z,\mathbf{r}}= \gamma_w$ for interbilayer junctions. The Langevin noise terms $\xi_{\mathbf{r}}$ and $\boldsymbol{\eta}_{\mathbf{r}}$ have a white Gaussian distribution with zero mean.
To satisfy the fluctuation-dissipation theorem, we take the noise of the order parameter as
\begin{align}
	\langle \mathrm{Re} \{{\xi_{\mathbf{r}} (t)}\} \mathrm{Re} \{{\xi_{\mathbf{r'}} (t')}\} \rangle &= \frac{\gamma_{\mathrm{sc}} k_{\mathrm{B}} T}{K \hbar^2 V_0} \delta_{\mathbf{r}\mathbf{r'}} \delta(t-t') \, , \\
	\langle \mathrm{Im} \{{\xi_{\mathbf{r}} (t)}\} \mathrm{Im} \{{\xi_{\mathbf{r'}} (t')}\} \rangle &= \frac{\gamma_{\mathrm{sc}} k_{\mathrm{B}} T}{K \hbar^2 V_0} \delta_{\mathbf{r}\mathbf{r'}} \delta(t-t') \, , \\
	\langle \mathrm{Re} \{{\xi_{\mathbf{r}} (t)}\} \mathrm{Im} \{{\xi_{\mathbf{r'}} (t')}\} \rangle &= 0 ,
\end{align}
where $V_0= d_{ab}^2 d_c$. The noise correlations for the vector potential are
\begin{align}
	\langle \eta_{x,\mathbf{r}} (t) \eta_{x,\mathbf{r'}} (t') \rangle &= \frac{2 \gamma_{ab} k_{\mathrm{B}} T}{\epsilon_\infty \epsilon_0 V_0} \delta_{\mathbf{r}\mathbf{r'}} \delta(t-t') , \\
	\langle \eta_{y,\mathbf{r}} (t) \eta_{y,\mathbf{r'}} (t') \rangle &= \frac{2 \gamma_{ab} k_{\mathrm{B}} T}{\epsilon_\infty \epsilon_0 V_0} \delta_{\mathbf{r}\mathbf{r'}} \delta(t-t') , \\
	\langle \eta_{z,\mathbf{r}} (t) \eta_{z,\mathbf{r'}} (t') \rangle &= \frac{2 \gamma_{z,\mathbf{r}} k_{\mathrm{B}} T}{\kappa_{z,\mathbf{r}} \epsilon_\infty \epsilon_0 V_0} \delta_{\mathbf{r}\mathbf{r'}} \delta(t-t') .
\end{align}

\section{Model parameters}
We simulate a bilayer cuprate with $N= 40 \times 40 \times 4$ lattice sites. The model parameters are specified in Table~\ref{tab:parameters}.

\begin{table}[!h]
	\centering
	\caption{Model parameters of the simulated bilayer cuprate.}
	\renewcommand{\arraystretch}{1.5}
	\begin{tabular}{lr}
		\hline
		$K~(\text{meV}^{-1})$ & $2.9 \times 10^{-5}$ \\
		$\mu~(\text{meV})$ & $1.0 \times 10^{-2}$ \\
		$g~(\text{meV} \, \text{\AA}^3)$ & 5.0 \\
		\hline
		$\epsilon_\infty$ & 4 \\
		$d_{ab}~(\text{\AA})$ & 15 \\
		$d_s~(\text{\AA})$ & 4 \\
		$d_w~(\text{\AA})$ & 8 \\
		$t_{ab}~(\text{meV})$ & $5.7 \times 10^{-1}$ \\
		$t_s~(\text{meV})$ & $\qquad 3.9 \times 10^{-2}$ \\
		$t_w~(\text{meV})$ & $\qquad 3.6 \times 10^{-4}$ \\
		\hline
		$\gamma_{\mathrm{H}}/2\pi~(\mathrm{THz})$ & 1.0 \\
		$\gamma_{ab}/2\pi~(\mathrm{THz})$ & 7.0 \\
		$\gamma_s/2\pi~(\mathrm{THz})$ & 1.2 \\
		$\gamma_w/2\pi~(\mathrm{THz})$ & 0.4 \\
		\hline
	\end{tabular}
	\renewcommand{\arraystretch}{1}
	\label{tab:parameters}
\end{table}

\section{Plasma resonances}
A bilayer superconductor has two longitudinal Josephson plasma modes, which govern the dynamics of the Cooper pairs along the $c$ axis. The ground state expressions \cite{VanDerMarel2001, Koyama2002} for the two Josephson plasma frequencies are
\begin{widetext}
	\begin{equation}
		\omega_{\mathrm{J1,J2}}^2 = \biggl( \frac{1}{2}+ \alpha_{s} \biggr) \Omega_{s}^2 + \biggl( \frac{1}{2}+ \alpha_{w} \biggr) \Omega_{w}^2 \mp \sqrt{ \biggl[\biggl( \frac{1}{2}+ \alpha_{s} \biggr) \Omega_{s}^2 - \biggl( \frac{1}{2}+ \alpha_{w} \biggr) \Omega_{w}^2 \biggr]^2 + 4 \alpha_{s} \alpha_{w} \Omega_{s}^2 \Omega_{w}^2 } .
	\end{equation}
\end{widetext}
The bare plasma frequencies of the interlayer junctions are given by
\begin{equation}
	\Omega_{s,w}= \sqrt{\frac{8 t_{s,w} |\psi_0|^2  e^2 d_c d_{s,w}}{\epsilon_\infty \epsilon_0 \hbar^2}} ,
\end{equation}
and the capacitive coupling constants \cite{Koyama1996, Machida1999, Machida2004} are
\begin{equation}
	\alpha_{s,w}= \frac{\epsilon_\infty \epsilon_0}{8 K |\psi_0|^2 e^2 d_c d_{s,w}} ,
\end{equation}
where $d_c= (d_s + d_w)/2$ is the average $c$-axis spacing. 
The parameter choice in Table~\ref{tab:parameters} implies $\alpha_w \approx 1$, and $\alpha_s \approx 2$. The bare plasma frequencies are $\Omega_w/2\pi \approx 0.9~\mathrm{THz}$ and $\Omega_s/2\pi \approx 6.3~\mathrm{THz}$, resulting in the Josephson plasma frequencies
\begin{align}
	\omega_{\mathrm{J1}} &\approx 2\pi \times 1.0~\mathrm{THz} , \\
	\omega_{\mathrm{J2}} &\approx \Omega_s \sqrt{1+ 2\alpha_s} \approx 2\pi \times 14.1~\mathrm{THz} .
\end{align}

\begin{figure}[!t]
	\centering
	\includegraphics[scale=1]{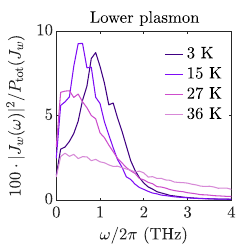}\llap{\parbox[b]{8cm}{{(a)}\\\rule{0ex}{3.7cm}}}
	\includegraphics[scale=1]{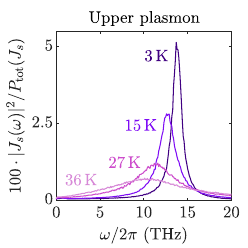}\llap{\parbox[b]{8cm}{{(b)}\\\rule{0ex}{3.7cm}}}
	\caption{Thermal distribution of the interlayer currents. (a) Power spectrum of the interbilayer current at different temperatures. (b) Power spectrum of the intrabilayer current at different temperatures. Each spectrum is based on an ensemble average of 1000 trajectories, and the spectral power is normalized by the total power. The crossover temperature is $T_c \sim 25~\mathrm{K}$.}
	\label{fig:fig2}
\end{figure}

To study the temperature dependence of the Josephson plasma modes, we compute the power spectra of the interlayer supercurrents based on an ensemble of 1000 trajectories.
The Josephson current along a single junction in the $z$ direction is given by
\begin{equation}
	J_{l,m,n}^z = \frac{2e t_{z,\mathbf{r}} d_c}{i \hbar} \left( \psi_{l,m,n+1}^* \psi_{l,m,n} e^{ia_{l,m,n}^z} - c.c. \right) .
\end{equation}
For each trajectory, we record the average interbilayer supercurrent
\begin{equation}
	J_w= \frac{1}{N_{xy}}\sum_{l,m} J_{l,m,2}^z
\end{equation}
and the average intrabilayer supercurrent
\begin{equation}
	J_s= \frac{1}{N_{xy}}\sum_{l,m} J_{l,m,1}^z
\end{equation}
over a time interval of 10~ps. We then compute the Fourier transforms $J_{w,s} (\omega)$ and evaluate the ensemble averages $\langle |J_{w,s} (\omega)|^2 \rangle$.
One can see in Fig.~\ref{fig:fig2}(a) that the lower Josephson plasma resonance shifts to lower frequencies with increasing temperature and vanishes around $T_c$. This is also observed in experiments \cite{Shibata1998, VonHoegen2022}.

\begin{figure}[!b]
	\centering
	\includegraphics[scale=1]{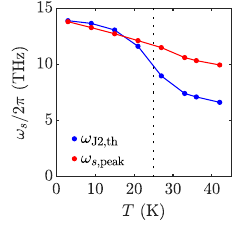}\llap{\parbox[b]{8cm}{{(a)}\\\rule{0ex}{3.5cm}}}
	\includegraphics[scale=1]{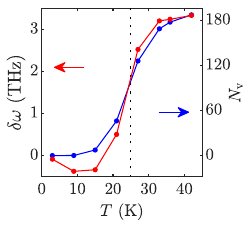}\llap{\parbox[b]{8cm}{{(b)}\\\rule{0ex}{3.5cm}}}
	\caption{Temperature dependence of the upper Josephson plasma resonance. (a) Temperature dependence of the peak frequency and the thermal average of the upper Josephson plasma frequency. (b) The discrepancy $\delta \omega= \omega_{s,\mathrm{peak}} - \omega_{\mathrm{J2,th}}$ has a similar temperature dependence as the number of vortices per layer $N_{\mathrm{v}}$. The crossover temperature is $T_c \sim 25~\mathrm{K}$.}
	\label{fig:fig3}
\end{figure}

As shown in Fig.~\ref{fig:fig2}(b), the upper Josephson plasma resonance broadens with increasing temperature and shifts to lower frequencies. The latter is also visible in Fig.~\ref{fig:fig3}(a), where the peak frequency $\omega_{s,\mathrm{peak}}$ is displayed as a function of temperature.
We compare the peak frequency to the thermal average
\begin{equation}
	\omega_{\mathrm{J2,th}} = \Omega_s \, \sqrt{\frac{\langle C_{l,m}^s \rangle}{|\psi_0|^2} + \alpha_s \left\langle \frac{C_{l,m}^s}{|\psi_{l,m,1}|^2} + \frac{C_{l,m}^s}{|\psi_{l,m,2}|^2} \right\rangle} ,
\end{equation}
where
\begin{equation}
	C_{l,m}^s= \frac{1}{2} \left( \psi_{l,m,2}^* \psi_{l,m,1} e^{ia_{l,m,1}^z} + c.c. \right) .
\end{equation}
This simple estimate describes the average renormalization of the plasma frequency of the intrabilayer junctions. It accounts for the renormalization of both the effective intrabilayer tunneling and the order parameter due to thermal fluctuations. At temperatures above 20~K, $\omega_{\mathrm{J2,th}}$ is clearly smaller than $\omega_{s,\mathrm{peak}}$. Remarkably, the discrepancy $\delta \omega= \omega_{s,\mathrm{peak}} - \omega_{\mathrm{J2,th}}$ follows a similar temperature dependence as the areal density of vortices, which is highlighted by Fig.~\ref{fig:fig3}(b). This indicates that the appearance of vortices leads to a significant stabilization of the upper Josephson plasma frequency.

\begin{figure}[!t]
	\centering
	\includegraphics[scale=1]{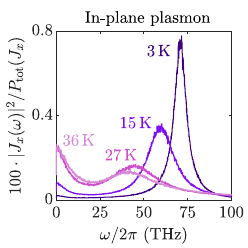}\llap{\parbox[b]{8cm}{{(a)}\\\rule{0ex}{3.7cm}}}
	\includegraphics[scale=1]{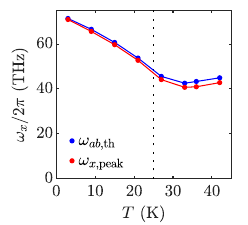}\llap{\parbox[b]{8cm}{{(b)}\\\rule{0ex}{3.7cm}}}
	\caption{Temperature dependence of the in-plane plasma resonance. (a) Power spectrum of the in-plane currents at different temperatures. Each spectrum is based on an ensemble average of 1000 trajectories, and the spectral power is normalized by the total power. (b) Temperature dependence of the peak frequency and the thermal average of the in-plane plasma frequency. The crossover temperature is $T_c \sim 25~\mathrm{K}$.}
	\label{fig:fig4}
\end{figure}

We now turn to the temperature dependence of the in-plane plasma frequency. At zero temperature, the in-plane plasma frequency is
\begin{equation}
	\omega_{ab}= \sqrt{\frac{8 t_{ab} |\psi_0|^2 e^2 d_{ab}^2}{\epsilon_\infty \epsilon_0 \hbar^2}} \approx 2\pi \times 73.7~\mathrm{THz} .
\end{equation}
We evaluate the average supercurrent along the $x$~axis based on the definition
\begin{equation}
	J_x= \frac{1}{2N_{xy}}\sum_{l,m} \sum_{n=1}^2 J_{l,m,n}^x ,
\end{equation}
where
\begin{equation}
	J_{l,m,n}^x = \frac{2e t_{ab} d_{ab}}{i \hbar} \left( \psi_{l+1,m,n}^* \psi_{l,m,n} e^{ia_{l,m,n}^x} - c.c. \right) .
\end{equation}
In Fig.~\ref{fig:fig4}(a), we present the power spectrum of the supercurrents along the $x$~axis at different temperatures. Similarly to the upper Josephson plasma resonance, the in-plane plasma resonance broadens with increasing temperature. While the peak frequency $\omega_{x,\mathrm{peak}}$ decreases monotonically with increasing temperature below $T_c$, it slowly increases above $T_c$. This behavior is consistent with the temperature dependence of the order parameter. Indeed, we find that the temperature dependence of $\omega_{x,\mathrm{peak}}$ is described by the average renormalization of the plasma frequency of the in-plane junctions. As evidenced by Fig.~\ref{fig:fig4}(b), $\omega_{x,\mathrm{peak}}$ is in good agreement with the thermal average
\begin{equation}
	\omega_{ab,\mathrm{th}} = \omega_{ab} \, \sqrt{ \frac{\langle \psi_{l+1,m,n}^* \psi_{l,m,n} e^{ia_{l,m,n}^x} + c.c. \rangle}{2|\psi_0|^2}}
\end{equation}
at all simulated temperatures.

\section{Temperature dependence of the in-plane tunneling and the superconducting order parameter}
In this section, we show the behavior of the effective in-plane tunneling coefficient and the amplitude of the order parameter across the phase ordering transition described in the main text.

\begin{figure}[!t]
	\centering
	\includegraphics[scale=1]{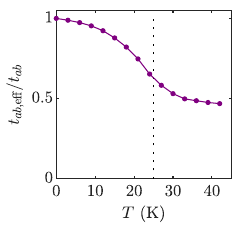}\llap{\parbox[b]{8cm}{{(a)}\\\rule{0ex}{3.4cm}}}
	\includegraphics[scale=1]{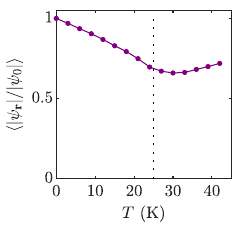}\llap{\parbox[b]{8cm}{{(b)}\\\rule{0ex}{3.4cm}}}
	\caption{Temperature dependence of the in-plane tunneling and the superconducting order parameter. (a) Temperature dependence of the effective in-plane tunneling coefficient. (b) Temperature dependence of the amplitude of the order parameter. Both quantities are averaged over all lattice sites, a time interval of 2~ps (200 measurements), and an ensemble of 100 trajectories. The standard error of each data point is comparable to the point size. The crossover temperature is $T_c \sim 25~\mathrm{K}$.}
	\label{fig:fig1}
\end{figure}

The temperature dependence of the effective in-plane tunneling coefficient $t_{ab, \mathrm{eff}} = t_{ab} \left\langle \cos \theta_{\mathbf{r}}^x \right\rangle$ is plotted in Fig.~\ref{fig:fig1}(a). Local phase fluctuations reduce the effective in-plane tunneling coefficient through disorder averaging of $\left\langle \cos \theta_{\mathbf{r}}^x \right\rangle$.

The temperature dependence of the order parameter is displayed in Fig.~\ref{fig:fig1}(b). In the ground state at $T=0$, the amplitude of the order parameter is given by $|\psi_0|^2= \mu/g= 2 \times 10^{21}~\mathrm{cm}^{-3}$.
The order parameter first decreases with increasing temperature and reaches a minimum of $\langle |\psi_{\mathbf{r}}| \rangle/|\psi_0| \approx 0.66$ around 30~K. Above 30~K, it slowly increases with increasing temperature. This behavior is consistent with an order-disorder transition \cite{Schneider1976}. We attribute the temperature dependence of the order parameter to a modification of the order parameter potential due to phase fluctuations, depleting the order parameter at temperatures below 30~K. However, as temperature is further increased, amplitude fluctuations are also strongly excited, leading to an increase of $\langle |\psi_{\mathbf{r}}| \rangle$.

\section{Vortex correlations}
\begin{figure*}[!t]
	\centering
	\includegraphics[scale=1]{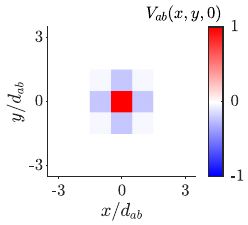}\llap{\parbox[b]{8cm}{{(a)}\\\rule{0ex}{3.4cm}}}
	\includegraphics[scale=1]{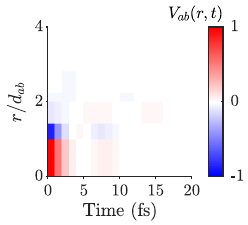}\llap{\parbox[b]{8cm}{{(b)}\\\rule{0ex}{3.4cm}}}
	\includegraphics[scale=1]{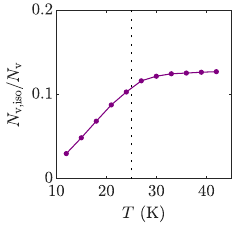}\llap{\parbox[b]{8.15cm}{{(c)}\\\rule{0ex}{3.4cm}}}
	\includegraphics[scale=1]{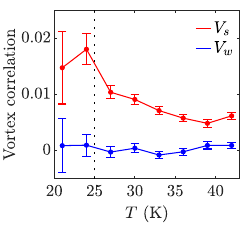}\llap{\parbox[b]{8.15cm}{{(d)}\\\rule{0ex}{3.4cm}}}
	\caption{Vortex excitations. (a) Equal-time in-plane vortex correlation function at $36~\mathrm{K} \sim 1.4 T_c$. (b) Time-resolved in-plane vortex correlation function at 36~K, where $r^2= x^2 + y^2$. (c) Relative amount of isolated vortices at different temperatures. Isolated vortices are vortices without an antivortex on the nearest or next-nearest neighbor plaquettes. (d) Interlayer vortex correlations. The results in (a) and (b) are obtained from an ensemble average of 1000 trajectories. Each data point in (c) and (d) is based on an ensemble average of 100 trajectories.}
	\label{fig:fig5}
\end{figure*}

Using the definition for the vorticity given by Eq.~(4) in the main text, we define the two-point correlation function of vortices in the $xy$ plane,
\begin{equation}
	V_{ab} (x_i,y_j,t)= \frac{\langle v_{l,m,n} (0) v_{l+i,m+j,n} (t) \rangle}{\langle v_{l,m,n}^2 (0) \rangle} .
\end{equation}
In Fig.~\ref{fig:fig5}(a), we show the equal-time in-plane vortex correlation function at 36~K. It reveals a strong tendency to the formation of vortex-antivortex pairs. The accumulated probability to find an antivortex on the nearest or next-nearest plaquettes of a vortex amounts to $94\%$. On larger length scales, in-plane vortex correlations are negligible.
Next, we consider the cumulative correlation function
\begin{equation}
	V_{ab} (r,t)= \sum_{|(x_i,y_j)|=r} V_{ab} (x_i,y_j,t) ,
\end{equation}
where the sum is taken over all $(x_i, y_j)$ with $x_i^2 + y_j^2= r^2$. One can see in Fig.~\ref{fig:fig5}(b) that vortex-antivortex pairs annihilate on a time scale of a few femtoseconds.

In Fig.~\ref{fig:fig5}(c), we show the relative amount of isolated vortices as a function of temperature. An isolated vortex is a vortex without an vortex of opposite vorticity on the nearest or next-nearest neighbor plaquettes. The percentage of isolated vortices grows below $T_c$ and saturates at higher temperature. This indicates a transition from bound to unbound vortices akin to a Kosterlitz-Thouless transition. The percentage of isolated vortices is limited by the areal density of vortices.

Furthermore, we calculate the interlayer correlation functions
\begin{align}
	V_s &= \frac{\langle \tilde{v}_{l,m,1} \tilde{v}_{l,m,2} \rangle}{\langle \tilde{v}_{l,m,n}^2 \rangle} , \\
	V_w &= \frac{\langle \tilde{v}_{l,m,2} \tilde{v}_{l,m,3} \rangle}{\langle \tilde{v}_{l,m,n}^2 \rangle} ,
\end{align}
where
\begin{equation}
	\tilde{v}_{l,m,n}= \sum_{l'=l-1}^{l+1} \sum_{m'=m-1}^{m+1} v_{l',m',n}
\end{equation}
is the vorticity of a bin of 9 plaquettes. The interlayer vortex correlation functions are displayed in Fig.~\ref{fig:fig5}(d). While interbilayer vortex correlations are generally negligible, intrabilayer vortex correlations are larger than zero at all simulated temperatures. The intrabilayer correlations are small and follow a similar temperature dependence as the effective intrabilayer coupling; see Fig.~2(b) in the main text. Note that the in-plane penetration depth $\lambda_{ab}= \omega_{ab}/c\sqrt{\epsilon_\infty} \approx 324~\mathrm{nm}$ is larger than the in-plane system size of 60~nm.

\section{Correlations of the intrabilayer Josephson potential} 

\begin{figure}[!h]
	\centering
	\includegraphics[scale=1]{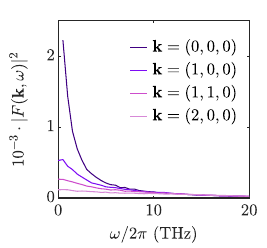}\llap{\parbox[b]{8.5cm}{{(a)}\\\rule{0ex}{3.7cm}}}
	\includegraphics[scale=1]{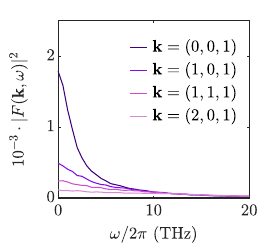}\llap{\parbox[b]{8.5cm}{{(b)}\\\rule{0ex}{3.7cm}}}
	\caption{Power spectra of the disorder function at $36~\mathrm{K} \sim 1.4 T_c$. (a) Power spectra of the disorder function for various momenta with $k_z=0$. (b) Power spectra of the disorder function for various momenta with $k_z=\pi/2d_c$. The in-plane momentum components are specified in units of $2\pi/L_{ab}$. The power spectra are evaluated from an ensemble average of 1000 trajectories.}
	\label{fig:fig6}
\end{figure}

In the presence of fluctuating vortices, the intrabilayer Josephson potential becomes disordered and fluctuating. We characterize this effect by computing the power spectra of spatial and time variations around the spatiotemporal mean, through the function
\begin{equation}
    F(\mathbf{r},t)= \cos \theta_s (\mathbf{r},t) - \langle \cos \theta_s (\mathbf{r},t) \rangle .
\end{equation}
For each trajectory of an ensemble of 1000 trajectories, we record $F(\mathbf{r},t)$ for 2~ps with a detection rate of 5~PHz and compute the Fourier transform
\begin{equation}
	F(\mathbf{k}, \omega)=  \frac{1}{N_t} \sum_{\mathbf{r}} \sum_j F(\mathbf{r},t_j) \, e^{i(\mathbf{k} \cdot \mathbf{r} - \omega t_j)} ,
\end{equation}
where $N_t= 10^4$ is the number of measurements per trajectory.
In Fig.~\ref{fig:fig6}, we show a selection of power spectra $|F(\mathbf{k},\omega)|^2$, based on the ensemble average of 1000 trajectories. Note that $F(\mathbf{k}=0, \omega=0)= 0$. We find that the disordered potential is peaked at the lowest momenta. As a function of frequency, we see noisy dynamics with fluctuations up to $\sim 5$ THz.

\newpage
\section{Details on the conductivity measurements}
\begin{figure}[!b]
	\centering
	\includegraphics[scale=1]{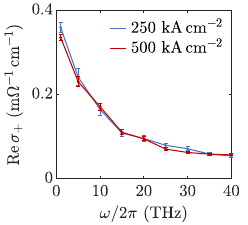}\llap{\parbox[b]{8.15cm}{{(a)}\\\rule{0ex}{3.45cm}}}
	\includegraphics[scale=1]{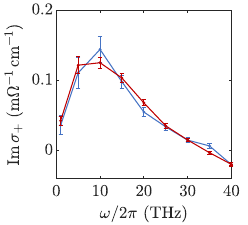}\llap{\parbox[b]{8.15cm}{{(b)}\\\rule{0ex}{3.45cm}}}
	\caption{Symmetric conductivity for different probe strengths at $36~\mathrm{K} \sim 1.4 T_c$. (a) Real part. (b) Imaginary part. The error bars indicate the standard errors of the ensemble averages.}
	\label{fig:fig7}
\end{figure}

In the main text, we present numerical results for the symmetric and antisymmetric components of the in-plane conductivity.
To measure the symmetric conductivity $\sigma_+$, we add a spatially homogeneous probe current to the equations of motion for $A_{x,\mathbf{r}}$,
\begin{equation}
	\partial_t^2 A_{l,m,n}^x = \frac{1}{\epsilon_\infty \epsilon_0} \frac{\partial \mathcal{L}}{\partial A_{l,m,n}^x} - \gamma_{ab} \partial_t A_{l,m,n}^x + \eta_{l,m,n}^x - \frac{J_{\mathrm{sym}}}{\epsilon_\infty \epsilon_0} \cos(\omega_{\mathrm{pr}} t) .
\end{equation}
Once a steady state is reached, we record the symmetric component of the electric field
\begin{equation}
	E_+= \frac{1}{N} \sum_{l,m,n} E_{l,m,n}^x=  \frac{1}{N} \sum_{l,m,n} \left(- \partial_t A_{l,m,n}^x \right)
\end{equation}
and the symmetric component of the current 
\begin{equation}
	J_+= \frac{1}{N_z} \sum_{n} J_n^x .
\end{equation}
The average current $J_n^x$ in layer $n$ includes superconducting, normal and capacitive contributions,
\begin{equation}
	J_n^x=  J_{n,\mathrm{sup}}^x + J_{n,\mathrm{nor}}^x + J_{n,\mathrm{cap}}^x .
\end{equation}
The superconducting current is given by
\begin{equation}
	J_{n,\mathrm{sup}}^x = \frac{1}{N_{xy}} \sum_{l,m} \frac{2e t_{ab} d_{ab}}{i \hbar} \left(\psi_{l+1,m,n}^* \psi_{l,m,n} e^{i a_{l,m,n}^x} - c.c. \right) .
\end{equation}
The normal current is given by
\begin{equation}
	J_{n,\mathrm{nor}}^x = \frac{1}{N_{xy}} \sum_{l,m} \epsilon_\infty \epsilon_0 \gamma_{ab} E_{l,m,n}^x .
\end{equation}
The capacitive current is given by
\begin{equation}
	J_{n,\mathrm{cap}}^x = \frac{1}{N_{xy}} \sum_{l,m} \epsilon_\infty \epsilon_0 \partial_t E_{l,m,n}^x .
\end{equation}
For $\omega_{\mathrm{pr}}/2\pi= 1~\mathrm{THz}$, we record $E_+ (t)$ and $J_+ (t)$ for 20~ps. For all other probe frequencies, we record $E_+ (t)$ and $J_+ (t)$ for 4~ps. Following this protocol, we evaluate $\sigma_+ (\omega_{\mathrm{pr}})= J_+(\omega_{\mathrm{pr}})/E_+(\omega_{\mathrm{pr}})$ for 100--1000 trajectories and take the ensemble average. We use $J_{\mathrm{sym}}= 500~\mathrm{kA \, cm^{-2}}$. Thus, we probe the linear response as evidenced by Fig.~\ref{fig:fig7}.

\begin{figure}[!t]
	\centering
	\includegraphics[scale=1]{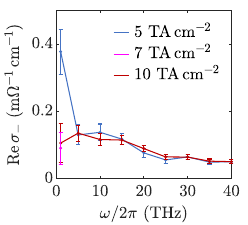}\llap{\parbox[b]{8.15cm}{{(a)}\\\rule{0ex}{3.45cm}}}
	\includegraphics[scale=1]{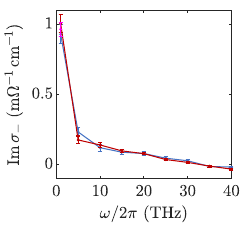}\llap{\parbox[b]{8.15cm}{{(b)}\\\rule{0ex}{3.45cm}}}
	\caption{Antisymmetric conductivity for different probe strengths at $36~\mathrm{K} \sim 1.4 T_c$. (a) Real part. (b) Imaginary part. The error bars indicate the standard errors of the ensemble averages.}
	\label{fig:fig8}
\end{figure}

To measure the antisymmetric conductivity $\sigma_-$, we proceed analogously to the symmetric case. Here, the probe current alternates from layer to layer, i.e.,
\begin{equation}
	\partial_t^2 A_{l,m,n}^x = \frac{1}{\epsilon_\infty \epsilon_0} \frac{\partial \mathcal{L}}{\partial A_{l,m,n}^x} - \gamma_{ab} \partial_t A_{l,m,n}^x + \eta_{l,m,n}^x - \frac{(-1)^n J_{\mathrm{asym}}}{\epsilon_\infty \epsilon_0} \cos(\omega_{\mathrm{pr}} t) .
\end{equation}
Once a steady state is reached, we record the antisymmetric component of the electric field
\begin{equation}
	E_-= \frac{1}{N} \sum_{l,m,n} (-1)^n E_{l,m,n}^x
\end{equation}
and the antisymmetric component of the current 
\begin{equation}
	J_-= \frac{1}{N_z} \sum_{n} (-1)^n J_n^x .
\end{equation}
For $\omega_{\mathrm{pr}}/2\pi= 1~\mathrm{THz}$, we record $E_+ (t)$ and $J_+ (t)$ for 20~ps. For all other probe frequencies, we record $E_+ (t)$ and $J_+ (t)$ for 4~ps. Following this protocol, we evaluate $\sigma_- (\omega_{\mathrm{pr}})= J_-(\omega_{\mathrm{pr}})/E_-(\omega_{\mathrm{pr}})$ for 100--1000 trajectories and take the ensemble average. In Fig.~\ref{fig:fig8}, we show data for probe strengths of $J_{\mathrm{asym}}= 5 \times 10^9~\mathrm{kA \, cm^{-2}}$, $J_{\mathrm{asym}}= 7 \times 10^9~\mathrm{kA \, cm^{-2}}$, and $J_{\mathrm{asym}}= 10^{10}~\mathrm{kA \, cm^{-2}}$. We suspect that the measurement with $J_{\mathrm{asym}}= 5 \times 10^9~\mathrm{kA \, cm^{-2}}$ at 1~THz is poorly converged and use the data from the measurement with $J_{\mathrm{asym}}= 7 \times 10^9~\mathrm{kA \, cm^{-2}}$ in the main text. For all other probe strengths, we use the results from the measurements with $J_{\mathrm{asym}}= 5 \times 10^9~\mathrm{kA \, cm^{-2}}$. Note that $J_{\mathrm{asym}}$ is significantly larger than $J_{\mathrm{sym}}$ because the antisymmetric probe current induces a magnetic field that strongly screens the antisymmetric current.

\begin{figure*}[!h]
	\centering
	\includegraphics[scale=1]{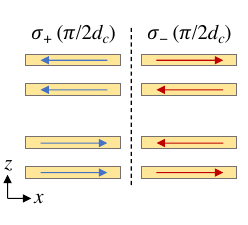}\llap{\parbox[b]{8.15cm}{{(a)}\\\rule{0ex}{3.4cm}}} \hspace{0.1cm}
	\includegraphics[scale=1]{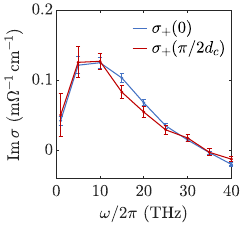}\llap{\parbox[b]{8.15cm}{{(b)}\\\rule{0ex}{3.4cm}}}
	\includegraphics[scale=1]{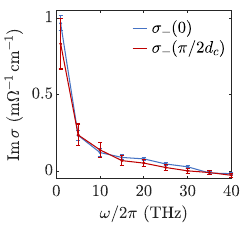}\llap{\parbox[b]{8.15cm}{{(c)}\\\rule{0ex}{3.4cm}}}
	\caption{Dependence of the symmetric and antisymmetric conductivity on the $z$-axis momentum. (a) Current configurations characterized by the symmetric and antisymmetric conductivity, respectively, for $k_z= \pi/2d_c$. (b) Imaginary part of $\sigma_+$ for $k_z=0$ and $k_z= \pi/2d_c$ at $36~\mathrm{K} \sim 1.4 T_c$. (c) Imaginary part of $\sigma_-$ for $k_z=0$ and $k_z= \pi/2d_c$ at 36~K. The error bars indicate the standard errors of the ensemble averages.}
	\label{fig:fig9}
\end{figure*}

Furthermore, we measure the symmetric and antisymmetric conductivity for a $z$-axis momentum of $k_z= \pi/2d_c$. The corresponding current configurations are depicted in Fig.~\ref{fig:fig9}(a). One can see in Figs.~\ref{fig:fig9}(b) and \ref{fig:fig9}(c) that $\sigma_+$ and $\sigma_-$ have no significant dependence on $k_z$. This confirms that the interbilayer coupling between vortices is negligible as indicated by Fig.~\ref{fig:fig5}(d).

\bibliography{biblio}